\documentclass[graybox]{svmult}

\usepackage{savesym}
\usepackage{amsmath}
\savesymbol{iint}
\usepackage{txfonts}
\restoresymbol{TXF}{iint}

\usepackage[normalem]{ulem}
\usepackage{mathptmx}       
\usepackage{helvet}         
\usepackage{courier}        

\usepackage{graphicx}
\usepackage[bottom]{footmisc}
\usepackage{hyperref}
\usepackage{soul}
\hypersetup{colorlinks=true,urlcolor=blue}

\usepackage[square,numbers,sort&compress]{natbib}

\makeindex             

\usepackage[utf8]{inputenc}

\def\bea{\begin{eqnarray}}
\def\ea{\end{eqnarray}}
\def\be{\begin{equation}}
\def\ee{\end{equation}}
\def\f{\frac}
\def\lp{\ell_{\rm Pl}}

\def\Q{\mathcal{Q}}

\def\h{\hat}
\def\f{\frac}

\def\vk{\vec{k}}
\def\pp{\mathfrak{p}}
\def\d{{\rm d}}
 
\begin{document}


\title*{Loop quantum cosmology: \\ relation between theory and observations}

\author{Ivan Agullo, Anzhong Wang and Edward Wilson-Ewing}

\institute{Ivan Agull\'o \at Department of Physics and Astronomy,
Louisiana State University, \\ Baton Rouge, Louisiana 70803, USA. \\ \email{agullo@lsu.edu}
\and Anzhong Wang \at GCAP-CASPER, Department of Physics,
Baylor University, \\ Waco, Texas 76798-7316, USA. \\ \email{Anzhong\_Wang@baylor.edu}
\and Edward Wilson-Ewing \at Department of Mathematics and Statistics,
University of New Brunswick, \\ Fredericton, NB, E3B 5A3, Canada. \\ \email{edward.wilson-ewing@unb.ca}}

\maketitle

\abstract{This chapter provides a review of the frameworks developed for cosmological perturbation theory in loop quantum cosmology, and applications to various models of the early universe including inflation, ekpyrosis and the matter bounce, with an emphasis on potential observational consequences. It also includes a discussion on extensions to include non-Gaussianities and background anisotropies, as well as on its limitations concerning trans-Planckian perturbations and quantization ambiguities. It concludes with a summary of recent work studying the relation between loop quantum cosmology and full loop quantum gravity.
}

\section*{Keywords} 
Loop quantum gravity, loop quantum cosmology, cosmological perturbation theory, cosmic microwave background, inflation.

\section{Introduction}
\label{intro}

Loop quantum cosmology (LQC) is a quantum theory for the gravitational field of homogeneous spacetimes commonly used in cosmology, that is based on the techniques of loop quantum gravity (LQG). As has been reviewed in the previous chapter of this book, in recent years LQC has led to significant insights and detailed results concerning the quantization of these cosmological models, and fundamental questions have been addressed---in particular, the classical big-bang singularity is resolved by a non-singular bounce due to quantum gravity effects. LQC thus provides a detailed description of the spacetime geometry during the Planck era for Friedman-Lema\^itre-Robertson-Walker (FLRW) and Bianchi spacetimes.

The natural next step is to use LQC to extend our current model of the early universe to include Planck scale physics, by developing a framework for cosmological perturbation theory in LQC. The goal of such an extension is two-fold. On the one hand, it would allow us to overcome the limitations of general relativity, on which the standard cosmological model rests, and to achieve a more complete picture of the past history of the cosmos. And, on the other hand, such an extension could potentially connect Planck-scale physics with observations---in particular of the cosmic microwave background (CMB), the faint afterglow of the primordial universe---thereby opening an avenue to test some of the ideas on which this approach to quantum gravity rests. Describing the state of the art of this research program in a pedagogical yet comprehensive way is the primary goal of this chapter.

An additional motivation to study cosmological perturbations is more conceptual. Homogeneous spacetimes have a finite number of degrees of freedom, while cosmological perturbations are described by fields with local degrees of freedom. Cosmology, in addition to offering the possibility of comparing predictions to observations, also offers a simple testing ground for various tools and techniques of full quantum gravity, and this testing ground will be vastly enriched by extending it to have local degrees of freedom.

Including local degrees of freedom  is a challenging task, as the simplifying consequences of exact homogeneity can no longer be used. A variety of different approaches to extend LQC to include cosmological perturbations have been developed during the last decade, each with some simplifying assumptions and some strengths and weaknesses.

Importantly, these various extensions of LQC can provide a quantum gravity extension to many types of cosmological models including inflation, ekpyrosis, and the matter bounce. For example, in the standard inflationary scenario, one normally starts the evolution far from the Planck era, when the curvature and energy density of matter fields in the universe is around twelve orders of magnitude below the Planck scale and quantum gravity effects are negligible.  Our ignorance about the earlier stages of cosmic evolution is encoded in the choice of initial conditions at the onset of inflation, for both the background homogeneous geometry and for cosmological perturbations, with the latter typically assumed to be in the so-called Bunch-Davies vacuum at the onset of inflation. This is a key, yet strong assumption. It is of considerable interest to extend this scenario backwards in time to include the Planck era, and to show that such initial conditions (or something close to them) can be derived as a result of the pre-inflationary dynamics when quantum gravity effects become important. Further, in typical inflationary models the background spacetime is classically singular, but this is cured in LQC where quantum gravity effects replace the big-bang singularity by a non-singular bounce. Similarly, the ekpyrotic and matter bounce scenarios both require a cosmic bounce, and LQC provides a natural mechanism for such bounce.

This chapter provides an overview of how LQC provides a quantum gravity completion of these cosmological scenarios, and highlights the extra features that this extension adds to observable quantities. These new effects, due to quantum gravity, are a window to the Planck era of the cosmos, and can be used to test the ideas discussed here by comparing the predictions to observations of the CMB. In addition, these results also provide insight on how standard quantum field theory can emerge from a background independent approach to quantum gravity, among other foundational questions for quantum cosmology.

The outline of this chapter is the following: Sec.~\ref{s.class} provides a brief review of standard cosmological perturbations. Then Sec.~\ref{s.perts-lqc} presents four different and complementary LQC-based frameworks for cosmological perturbation theory, and Sec.~\ref{s.app} describes the results of applying these frameworks to different cosmological models, including inflation and some alternatives to inflation. Extensions to include non-Gaussianities and background anisotropies are reviewed in Sec.~\ref{s.ext}, and some limitations are discussed in Sec.~\ref{s.lim}. Finally, we end with some comments on the link between LQC and LQG in Sec.~\ref{s.lqg}. We use natural units where $c=G=\hbar=1$.

\section{Standard cosmology: review and guidance for quantum gravity}
\label{s.class}

Before discussing quantum gravity effects on cosmological perturbations, it is useful to review standard cosmological perturbation theory based on quantum field theory on a fixed classical background; for a detailed introduction, see, e.g., \cite{Mukhanov:1990me}. In addition to setting the notation and pointing out some key results, this discussion will give some basic intuition about the dynamics of cosmological perturbations, and provide some hints as to where quantum gravity effects may be expected to arise---for pedagogical purposes, this last part of the discussion is qualitative in this section; the way this picture concretely emerges in LQC is presented later in this chapter.

Throughout this chapter, we mostly focus on scalar perturbations to shorten the discussion. We provide references to the relevant literature for the interested reader for details on tensor and vector modes.

For concreteness, we will consider cosmological perturbations on a spatially flat FLRW background geometry for the case that the matter content is a minimally coupled scalar field $\phi$ sourced by a potential $V(\phi)$. (Other matter fields and homogeneous background geometries are possible; see for example Sec.~\ref{anisotropies} for a summary on the extension to Bianchi~I).  Perturbations to the metric tensor and the scalar field, $\delta g_{ab}(\vec x,t)$ and $\delta \phi(\vec x,t)$, contain three physical degrees of freedom: a scalar mode due to matter, and two tensor modes of purely gravitational origin, corresponding to the two polarizations of gravitational waves. Vector perturbations can be ignored in this scenario, as they do not get excited by scalar matter. 

The scalar mode can be described by the field $\mathcal{Q}$ that is gauge-invariant in the sense that it is invariant under linear diffeomorphisms; note that $\mathcal{Q}$ is related to the familiar comoving curvature perturbation $\mathcal{R}$ and the Mukhanov-Sasaki variable $v$ by $\mathcal{Q} \equiv {z}/a\,\mathcal{R} = v/a$, where $z = a \dot\phi/H$ and, as usual, $a(t)$ is the scale factor of the background spacetime, $H(t) = \dot a / a$ is the Hubble rate, and a dot denotes a derivative with respect to the cosmic time.

The classical phase space of interest is therefore $\Gamma_{\rm phys}=\Gamma_{\rm hom} \otimes \Gamma_{\rm pert}$, where $\Gamma_{\rm hom}$ is the standard phase-space of FLRW geometry, made of the two canonical pairs $(a,\pi_{a}, \phi, p_{\phi})$, and $ \Gamma_{\rm pert}$ is the phase-space of scalar and tensor perturbations. $\Gamma_{\rm pert}$ is commonly called the ``reduced phase-space'' of the perturbations because it only includes the gauge-invariant degrees of freedom of $\delta g_{ab}(\vec x,t)$ and $\delta \phi(\vec x,t)$. 

Still in the classical theory, the dynamics is determined as follows. The background dynamics in $\Gamma_{\rm hom}$ is simply determined by the usual Friedman equations for FLRW spacetimes, while the dynamics of the scalar is dictated by the true Hamiltonian (in Fourier space)
\be \label{pert-hams}
\mathcal{C}_2^{(\Q)}[N]= \f{N}{2 (2\pi)^3 }\,\, \int d^3 k  \, \left( \f{1}{a^{3}}\, |\pp^{(\Q)}_{\vk}|^2 + a\, (k^2+\mathcal{U}) |\Q_{\vk}|^2 \right) \, ,
\ee
where $N$ is the lapse function. This expression is obtained by expanding the scalar constraint of general relativity up to second order. Note that the scalar mode behaves as a minimally coupled scalar field with an effective potential
\be \label{potU}
\mathcal{U}= a^2 \left[ V(\phi) \, r-2 V_{\phi}(\phi) \sqrt{r} + V_{\phi\phi}(\phi) \right],
\ee
where $V_{\phi}(\phi)$  and $V_{\phi \phi}(\phi)$ are the first and second derivatives of the scalar field potential $V(\phi)$ with respect to $\phi$, while $r \equiv 3 \kappa p_{\phi}^2/[p_{\phi}^2/2 + a^6 V(\phi)]$.
The equation of motion, expressed in terms of conformal time $\eta$, is
\be  \label{waveeqs}
{\Q}_{\vk}^{\prime\prime} + 2 \f{{a}^\prime}{a}\, {\Q}_{\vk}^\prime + \Big( k^2 +{\mathcal{U}}(\eta) \Big) \, {\Q}_{\vk} = 0 ~.
\ee
Here primes denote derivatives with respect to the conformal time, with $f' = a \dot f$. This is a second-order ordinary differential equation with coefficients given by the background quantities, such as the scale factor $a(\eta)$ and its derivative. The dynamics for tensor modes are very similar, except with no effective potential, $\mathcal{U}_{\rm tensor} = 0$.

It can be convenient to rewrite the equation of motion in terms of other variables describing the scalar mode, for example, for the Mukhanov-Sasaki variable $v = aQ$
\be
\label{ms-class}
v_k'' + \left( k^2 - \f{z''}{z} \right) v_k = 0,
\ee
and $z''/z = a''/a - {\cal U}$.

Therefore, in this perturbative approach one first solves the evolution for the background degrees of freedom $a(\eta)$ and $\phi(\eta)$, while ignoring perturbations; this fixes the background spacetime metric once and for all. Next, the solution for $a(\eta)$ and $\phi(\eta)$ is needed for Eq.~\eqref{waveeqs} whose solution, in turn, determines the dynamics of the linear perturbations. This completes the summary of the classical theory. 

In approaches to the early universe such as inflation, one further proceeds to quantize the perturbations, while leaving the background geometry classical. The homogeneous phase space $\Gamma_{\rm hom}$ and its dynamics are unmodified, but the classical phase space for the perturbations $\Gamma_{\rm pert}$ is replaced by a Hilbert space $\mathcal{H}_{\rm pert}$, and the quantum dynamics is determined by interpreting \eqref{waveeqs} as the Heisenberg equation for the operator $\hat{\mathcal{Q}}_{\vk}$.

Importantly, $z''/z$ introduces a scale into the dynamics as can be most clearly seen from Eq.~\eqref{ms-class}. In inflation, this length scale is the Hubble radius $r_H = 1/H$, but more generally $\sqrt{|z/z''|}$ gives a measure of the radius of curvature of the background spacetime. Given the importance of this lengthscale, it is natural to split the Fourier modes $v_k$ into two categories: those with a wavelength shorter than $\sqrt{|z/z''|}$, and those with a longer wavelength.

For short-wavelength modes ($k^2 \gg |z''/z|$), the term $|z''/z|$ in \eqref{ms-class} is unimportant; in plain words, short-wavelength modes do not ``feel'' the curvature of the background spacetime and evolve exactly as in the Minkowski space, while long-wavelength perturbations ($k^2 \ll |z''/z|$) do feel the spacetime curvature and evolve differently, with in this case the $k^2$ term in the equation of motion being negligible.

This simple fact of the physics underlying cosmological perturbations helps to understand where quantum gravity effects might arise. There are two obvious possible regimes: (i) for short-wavelength modes that have a wavelength comparable to the Planck length $\lp$, (ii) for long-wavelength modes at a time when the dynamics of the background spacetime are modified by quantum gravity effects (since this will in turn introduce corrections in $z''/z$ which impacts long-wavelength perturbations).

For the first possibility, modifications to the equations of motion for cosmological perturbations may arise for wavelengths $\lambda \sim \lp$. Recalling that short-wavelength modes evolve just as in the Minkowski space, quantum gravity effects for these modes are typically modeled as modifications to the dispersion relation for the perturbations with the equation of motion becoming $v_k'' + f(k)^2 v_k = 0$ where $f(k)$ depends only on $k$ and is independent of the scale factor $a(t)$ and other geometric degrees of freedom of the background spacetime, and $f(k) \to k$ in the limit $\lambda \gg \lp$. It has been shown that so long as $f(k)$ is real and the background dynamics is sufficiently smooth so the quantum state $|v_k\rangle$ evolves adiabatically, modified dispersion relations will not have any impact on predictions for the CMB \cite{Brandenberger:2000wr}.

This leaves the second possibility: the evolution of long-wavelength perturbations depends on the background spacetime, and if the dynamics of the background are modified due to quantum gravity effects, then these will in turn have an impact on long-wavelength perturbations, which can potentially show up in the CMB.

This expectation due to general heuristic arguments is in fact realized and made precise by ab initio calculations in LQC, as shall be reviewed below.

\section{Cosmological perturbations in LQC}
\label{s.perts-lqc}

Progress in understanding the very early universe has been possible due to a key observation of the CMB: the early universe was extraordinarily homogeneous and isotropic \cite{Planck:2019evm}, and departures from exact homogeneity are sufficiently small that they can be described by linear perturbation theory on a FLRW geometry. 

As reviewed in Sec.~\ref{s.class}, in standard cosmology the dynamics of the background and the perturbations are derived from the Einstein equations; the background is treated classically while the fields describing the perturbations are quantized using linear quantum field theory in curved spacetimes, neglecting back-reaction on the background spacetime.

It is not possible to follow the same procedure in LQG because the quantum analog of the Einstein equations  is not yet fully understood. Much of the work in LQC initially focused on homogeneous spacetimes, but now there has been considerable work to extend the formalism of LQC to include perturbations. Here we summarize several strategies that have been developed to study cosmological perturbations in LQC, emphasizing their strengths and weaknesses.

\subsection{Dressed metric approach}
\label{s.dressed}

The dressed metric approach to cosmological perturbation theory in LQC is based on a framework for quantum field theory on a quantum background spacetime \cite{Ashtekar:2009mb}. It follows the strategy of semi-classical cosmology: restrict attention, already in the classical theory, to FLRW geometries with linear, gauge-invariant perturbations, and then quantize this sector \cite{Agullo:2012sh, Agullo:2012fc}. This is a natural extension of LQC applied to homogeneous spacetimes, where homogeneity is imposed classically and it is only the phase space reduced to the homogeneous degrees of freedom that is quantized. The idea underlying the dressed metric approach is to now include linear perturbations in the phase space that is to be quantized.

This approach rests on three key assumptions.
(i) Gauge-invariant perturbations are defined at the classical level, this assumes that quantum corrections to the definition of gauge-invariant perturbations are subleading compared to other quantum gravity effects.
(ii) The back-reaction of perturbations on the homogeneous geometry is neglected, as is standard for linear perturbation theory; the consistency of this assumption is checked a posteriori by verifying that perturbations remain sufficiently small, as is also done in the semi-classical theory. (iii) The quantization is based on a hybrid approach, where the homogeneous degrees of freedom are quantized following LQC, while perturbations are handled by standard quantum field theory. This hybrid approach was first introduced for Gowdy cosmologies \cite{Martin-Benito:2008eza} and ignores the effects of potential LQC corrections to the equations of motion for the perturbations themselves, which is motivated by the fact that the energy-momentum of curvature perturbations always remains well below the Planck scale \cite{Agullo:2013ai}. 

In general terms, this approach follows a very similar strategy to the strategy reviewed in Sec.~\ref{s.class}, with the important difference that the background geometry is now also quantized: the classical phase space $\Gamma_{\rm hom}$ is replaced by the Hilbert space $\mathcal{H}_{LQC}$ of LQC for FLRW spacetimes, described in the Handbook's chapter on homogeneous LQC \cite{LQC-chapter}. The FLRW background is therefore no longer described by classical functions $a(\eta)$ and $\phi(\eta)$, but instead by a wave-function $\Psi_{\rm hom}(a, \phi)$. The challenge now is to determine the dynamics for the perturbations given the quantum background spacetime $\Psi_{\rm hom}(a, \phi)$.

In more detail, following the principles of LQG, the dynamics for perturbation in the dressed metric approach is derived starting from the Wheeler-de Witt equation $\hat H_{\rm tot} \, \Psi_{\rm tot} = 0$, where $\Psi_{\rm tot}(a, \phi,\Q_{\vk})$ is the quantum state including both background and scalar perturbation degrees of freedom. The Hamiltonian $\hat H_{\rm tot} = \hat H_{\rm hom} + \hat H_{\rm pert}$ contains a background term $\hat H_{\rm hom} =  (\hat \Theta - \partial_\phi^2)/2$ corresponding to the LQC Hamiltonian constraint for the homogeneous FLRW spacetime (as reviewed in 
the Handbook's chapter on homogeneous LQC \cite{LQC-chapter}), with $\hat\Theta$ being the gravitational part of $\hat H_{\rm hom}$ while $\hat H_{\rm pert}$ is the Hamiltonian for the scalar perturbations (for the extension to tensor modes, see \cite{Agullo:2012sh}).

After deparameterization and interpreting $\phi$ as time, $\hat H_{\rm tot} \, \Psi_{\rm tot} = 0$
can be written as a Schrodinger equation,
\be \label{dressed-wdw}
-i \, \partial_\phi \Psi_{\rm tot} =  \left| \, \hat  \Theta
- 2 \, \hat H_{\rm pert} \, \right|^{\frac{1}{2}} \, \Psi_{\rm tot}\, .
\ee
Perturbative methods can be used to solve this equation, following steps analogous to those used for the classical theory. First, it is assumed the background geometry is not affected by the perturbations, so the total wave function has a product form $\Psi_{\rm tot}(a, \phi, \Q_{\vk}) = \Psi_{\rm hom}(a,\phi) \otimes \Psi_{\rm pert}(a, \phi,\Q_{\vk})$. Second, the operator  $\hat \Theta$ is interpreted as the Hamiltonian of the  `heavy' degree of freedom and $\hat H_{\rm pert}$ as the Hamiltonian of the light degree of freedom; using this to expand the square root in \eqref{dressed-wdw} gives \cite{Ashtekar:2009mb}
\be
-i\, \left( (\partial_\phi \Psi_{\rm hom}) \otimes \Psi_{\rm pert} + \Psi_{\rm hom} \otimes ( \partial_\phi \Psi_{\rm pert}) \right) =
\, \sqrt{\hat \Theta} \Psi_{\rm hom} \otimes \Psi_{\rm pert} - \hat H_{\rm pert}
(\Psi_{\rm hom}\otimes \Psi_{\rm pert})\, . \nonumber
\ee
Note that $\hat \Theta$ does not act on perturbations, but $\hat H_{\rm pert}$ acts on both  background and perturbation states, as it contains background as well as perturbation operators. The third step is to choose $\Psi_{\rm hom}$ so it satisfies the background quantum equation $-i \partial_\phi \Psi_{\rm hom} =  \sqrt{\hat \Theta} \Psi_{\rm hom}$. This makes the first term in the left and right sides of the previous equation to cancell out, and the remaining quantum dynamics
\be
\Psi_{\rm hom}\otimes  (i\, \partial_\phi \Psi_{\rm pert})
= \hat H_{\rm pert} (\Psi_{\rm hom}\otimes \Psi_{\rm pert})\, ,
\ee
can be used to solve for $\Psi_{\rm pert}$. Since the left side of this last equation is proportional to $\Psi_{\rm hom}$, it follows that so is the right side. Due to this property, no information is lost when taking the inner product of this equation with $\Psi_{\rm hom}$, which gives
\be  \label{qftqst}
i\, \partial_\phi \Psi_{\rm pert} = \langle  \hat H_{\rm pert} \rangle\,  \Psi_{\rm pert}\, ,
\ee
where the expectation value is taken in the state $\Psi_{\rm hom}$.

Interestingly, the previous equation shows that, at leading order, the evolution of perturbations propagating on a quantum FLRW geometry $\Psi_{\rm hom}$ is described by replacing the  background operators in the Hamiltonian ${ \hat{\mathcal{C}}}^{(\Q)}_{2}$ by their expectation value in the state $\Psi_{\rm hom}$. This, in turn, allows us to write the equations of motion in the Heisenberg picture, producing \cite{Agullo:2013ai}
\be \label{Qeqn}
\hat{\Q}_{\vk}^{\prime\prime} + 2 \f{\tilde{a}^\prime}{\tilde{a}}\, \hat{\Q}_{\vk}^\prime + \Big( k^2 +\tilde{\mathcal{U}}_d(\tilde\eta) \Big) \, \hat{\Q}_{\vk} = 0 \, ,
\ee
where the effective conformal time $\tilde \eta$ is defined by its relation to the internal time $\phi$ 
\be \label{etat}
\d\tilde{\eta} := \tilde{a}^2(\phi)\, \langle
\hat{\Theta}^{-\f{1}{2}}\rangle \, \d\phi\, ,
\ee
and
\be \label{at}
\quad\quad \tilde{a} := \left(\f{\langle
\hat{\Theta}^{-\f{1}{4}}\, \hat{a}^4(\phi)\,
\hat{\Theta}^{-\f{1}{4}}\rangle}{\langle \hat{\Theta}^{-\f{1}{2}}\rangle}\right)^{1/4}\, ,
\qquad
\tilde{\mathcal{U}}_d = \f{\langle \hat{\Theta}^{-\f{1}{4}}\, \hat{a}^2 \, \hat{\mathcal{U}}\, \hat{a}^2  \hat{\Theta}^{-\f{1}{4}}\rangle}{\langle \hat{\Theta}^{-\f{1}{4}}\, \hat{a}^4 \, \hat{\Theta}^{-\f{1}{4}}\rangle} \, .
\ee
The presence of the operator $\hat{\Theta}$ in these expressions originates from the lapse function $N_{\phi}$ associated with evolution in internal time $\phi$; this lapse is implicitly included in the Hamiltonian $\hat H_{\rm pert}$ in Eqn.~\eqref{qftqst}; see \cite{Agullo:2013ai} for  details. Note that whenever there are factor ordering ambiguities, a symmetric order has been chosen. Another ambiguity concerns the form of $\h {\cal{U}}$. Before quantization, it is possible to use the classical Friedman equations to rewrite  ${\cal{U}}$ (for example, replacing $H^2$ by $3/\kappa\, \rho$, with $\rho$ the energy density of the scalar field); but since the Friedman equations are modified in LQC, rewritings of ${\cal U}$ of this type produce inequivalent expressions for the quantum operator $\h {\cal{U}}$. This ambiguity is intrinsic to quantum theories with constraints, and it will arise on several occasions throughout this chapter.

A key result is that \eqref{Qeqn} has  the same general form as the semi-classical equation \eqref{waveeqs} for $\hat{\Q}$. Although the framework is conceptually very different, with a background described by a quantum LQC state, at leading order in perturbations  the evolution (\ref{Qeqn}) is identical to a field theory on a smooth FLRW metric with the effective, quantum-corrected line element
$\d s^2
=\tilde{a}^2(\tilde\eta)\, (-\d\tilde\eta^2 +
\, \d\vec{x}^2)\, .$
Finally, in terms of the Mukhanov-Sasaki variable $v$, the dynamics are
\be \label{vd} \hat v_{\vk}''+\left(k^2-\frac{\tilde{a}''}{\tilde a}+\tilde{\mathcal{U}_d}\right)\,  \hat v_{\vk}=0\, . \ee

These results merit a brief discussion. In LQC, the background homogeneous geometry is quantum, and described by a wave function $\Psi_{\rm hom}$. There is no smooth geometry on which other fields propagate, and no a priori notion of light-cones. Instead, the boundaries of causal propagation emerge from the full quantum dynamics: the propagation of $\mathcal{Q}$ is exactly that of fields on a globally hyperbolic FLRW spacetime with metric tensor $\tilde g_{ab}$, and the equations of motion are generally covariant and locally Lorentz invariant. In other words, a quantum field theory on a FLRW geometry emerges from the quantum geometry $\Psi_{\rm tot}$.

The metric $\tilde g_{ab}$ is commonly called the effective dressed metric \cite{Ashtekar:2009mb}; beyond approximating the physical background geometry, it encodes all the information contained in $\Psi_{\rm hom}$ that the test field $\mathcal{Q}$ is sensitive to. This is what the adjective ``effective'' refers to. On the other hand,  $\tilde g_{ab}$ is also called ``dressed'' because it depends not only  on the mean value of  $\Psi_{\rm hom}$ but also on some of its quantum fluctuations.

At a practical level, to evolve perturbations on a quantum FLRW geometry $\Psi_{\rm hom}$, it is sufficient to compute the  components of $\tilde g_{ab}$ from $\Psi_{\rm hom}$, and then proceed exactly as in standard quantum field theory in curved spacetimes.

If the state $\Psi_{\rm hom}$ is sharply peaked on a classical trajectory at late times (i.e., if $\Delta a/\langle a\rangle \ll 1$ at late times, where $\Delta a$ is the quantum dispersion of $\hat a$), then the scale factor $\tilde a$ of the dressed metric reduces to a solution of the effective equations of LQC, discussed in the Handbook's chapter on homogeneous LQC \cite{LQC-chapter}. In this case, the dressed metric approach becomes very simple and is formally identical to the semi-classical theory of cosmological perturbations, except with the scale factor replaced by a solution of the effective equations of LQC. Since the effective equations become indistinguishable from the classical Friedmann equations far from the Planck regime, this implies that the quantum field theory in quantum spacetimes just described reduces to standard quantum field theory on classical spacetimes in that regime.

The computation of the dressed metric $\tilde g_{ab}$ can be formally extended to background quantum states $\Psi_{\rm hom}$ that are not necessarily sharply peaked, but if quantum fluctuations are large then there arise infrared divergences due to ``long tails'' in $\Psi_{\rm hom}$ \cite{Kaminski:2019tqo, Kaminski:2019qjn}. This is reminiscent of the infrared divergences that arise in the calculation of the S-matrix in QED. One can envisage mechanisms for mathematically taming these divergences, but a better physical understanding of them is needed first.

The perturbative approach reviewed here, as well as the hybrid quantization described next, have been criticized for neglecting possible quantum corrections to the definition of gauge invariant perturbations while including quantum corrections for the background degrees of freedom, which may entail a violation of general covariance \cite{Bojowald:2020xlw} (although note general covariance is broken in all perturbative approaches, particularly once backreaction is neglected). As explained in assumption (i) at the beginning of Sec.~\ref{s.dressed}, working with the classical gauge-invariant variables consists in assuming that quantum corrections to gauge transformations are subleading compared to the impact of modified background dynamics on the perturbations. It would be desirable though to test this assumption from the viewpoint of full LQG.

Finally, recall that an important assumption underlying this formalism is the absence of significant backreaction of the perturbations on the homogeneous geometry.
This assumption can be checked by comparing the expectation value of the renormalized energy and pressure of perturbations with the background contribution \cite{Agullo:2013ai, Agullo:2012fc} and also by going to the next-to-leading order in perturbations \cite{Agullo:2017eyh}. The first approach suffers from the standard ambiguities in defining the renormalized energy-momentum tensor in quantum field theory in curved spacetimes \cite{Wald:1995yp}; using adiabatic regularization it has been shown that backreaction is  negligible throughout the evolution \cite{Agullo:2013ai}. Similarly, following the second approach, the contribution of perturbations at next-to-leading order is negligible; see Sec.~\ref{nonG} for details.

\subsection{Hybrid quantization}

Similar to the dressed metric approach, the hybrid quantization also adopts the philosophy of combining a loop quantization of the homogeneous and isotropic universe with a Fock quantization of the inhomogeneous perturbations \cite{Fernandez-Mendez:2012poe, Fernandez-Mendez:2014raa, Martinez:2016hmn}. However, the strategy in the two approaches is different. In the dressed metric approach, the total Hamiltonian is first split in two parts, namely the constraint for the background and the quadratic Hamiltonian for perturbations, which are treated separately, analogous to what is done in standard cosmology. On the other hand, in the hybrid approach, the total Hamiltonian is truncated at quadratic order in perturbations, and the background and quadratic Hamiltonians of the truncated system are treated together as a constrained symplectic system, following \cite{PhysRevD.31.1777}. In principle, the latter provides a path to include some backreaction of perturbations \cite{ElizagaNavascues:2020uyf}, although a consistent treatment requires going to second order in perturbation theory (since quadratic backreaction from linear perturbations is of the same order as linear backreaction from second-order perturbations) and this has not been worked out explicitly yet. When backreaction is neglected, the two approaches are classically equivalent for linear cosmological perturbations. However, the concrete implementations of the two approaches put forward in the literature so far differ partially because of different choices in factor-ordering ambiguities and in the choice of the definition of the operator $\hat{\mathcal{U}}$ \cite{Gomar:2015oea, CastelloGomar:2016rjj}. 

In particular, when neglecting the backreaction, in the hybrid approach the linearized equation for scalar perturbations is \cite{ElizagaNavascues:2017avq},
\be
\label{hybrid1a}
 v_k'' + \left(k^2 - \frac{4 \pi G}{3} a^2 (\rho - 3P) + \tilde{\cal U}_h \right) v_k = 0,
\ee
where
\be \label{Hybrid2b} 
\tilde{\cal U}_h = 
a^2\, \left[ V_{\phi\phi}(\phi) + 48\pi G\,  V(\phi) -  \frac{48\pi G}{\rho} V^2(\phi) + \frac{6a'\phi'}{a^3\rho}V_{\phi}(\phi)\right].
\ee 
The functions $\rho$ and $P$ are the energy density and pressure of the homogeneous universe;
for a scalar field with a potential $V(\phi)$, we have
$\rho = (\phi'/a)^2/2 + V(\phi)$ and $P = \rho - 2 V(\phi)$.
The potential  $\tilde {\cal U}_h$ and its analog    in the dressed metric, $\tilde {\cal U}_d$, have the same classical limit; but they are different in LQC close to the bounce (and the same is true for $(4\pi G/3)a^2(\rho-3P)$ and $a''/a$). This is an example of the ambiguity in the choice for the potential ${\cal U}$ for the perturbations: different combinations are possible that give the correct classical limit.

It is interesting that the power spectra for the scalar and tensor perturbations calculated in these two approaches are quite similar \cite{Agullo:2013ai, Agullo:2015tca, Bonga:2015xna, Zhu:2017jew, CastelloGomar:2017kbo, Wu:2018sbr, ElizagaNavascues:2020uyf}, despite the fact that the effective potentials near the  bounce can be different \cite{ElizagaNavascues:2017avq, Iteanu:2022zha}. 
 
The similarity of the results for physical observables, in spite of these differences, is mainly because the current observational frequency band observed in the CMB today corresponds to  ultraviolet frequencies at the time of the  quantum bounce, for which the effects of the potential are negligibly small; this is true both for the hybrid \cite{Wu:2018sbr, Li:2021mop} and the dressed metric approaches \cite{Zhu:2017jew, Li:2021mop}. For further discussions on this point, see \cite{deBlas:2016puz, Li:2022evi} and Sec.~\ref{s.inf-app}.

\subsection{Separate universe loop quantization}
\label{s.sep}

Another approach to cosmological perturbation theory in LQC is based on an adaptation of the `separate universe' framework \cite{Salopek:1990jq, Wands:2000dp}. The basic idea underlying this approach is that ``derivative'' terms are negligible for long-wavelength perturbations. Specifically, for a Fourier mode whose wavelength is greater than the Hubble radius, $k^2 \ll z''/z$ in the equation of motion \eqref{ms-class} and therefore the $k^2$ term can safely be neglected.

Due to this simplification, it is possible to calculate LQC corrections for long-wavelength modes in a relatively direct manner \cite{Wilson-Ewing:2012dcf, Wilson-Ewing:2015sfx}. First, introduce a spatial discretization, such that the lattice spacing is greater than $\sqrt{|z/z''|}$; clearly this only captures long-wavelength perturbations. Note that neglecting the $k^2$ term in Fourier space implies neglecting derivatives in position space, and on a lattice this implies neglecting interaction terms between neighbouring cells in the lattice. Second, for each cell in the discretization the scale factor $a(\vec x)$ is uniform and therefore the spacetime geometry of each cell corresponds to a homogeneous spacetime. Working in the longitudinal gauge, the spatial metric has the form $q_{ab} = {\rm diag} (a(\vec x)^2, a(\vec x)^2, a(\vec x)^2)$ with $a(\vec x) = a\, (1-\psi(\vec x))$, which (for each cell) is exactly the flat FLRW metric. Of course, the scale factor $a(\vec x)$ and the energy density in each cell in the lattice cannot vary too much from one cell to another if the discretization is to describe small perturbations on a homogeneous background.
Finally, the dynamics in each cell are generated by precisely the Hamiltonian constraint of the flat FLRW spacetime (with no new terms since interactions are being neglected for the long-wavelength modes being studied here) so the standard loop quantization for FLRW (reviewed in the Handbook's chapter on homogeneous LQC \cite{LQC-chapter}) can be applied, without any modifications, to each cell. This process gives a loop quantization for long-wavelength scalar perturbations.

When quantum fluctuations are small, the dynamics of the long-wavelength scalar perturbations are well approximated by the semi-classical equation
\be
v_k'' - \f{z''}{z} v_k = 0\, ,
\ee
where $v$ is the Mukhanov-Sasaki variable introduced in \eqref{ms-class}. Note that this equation clearly has the correct classical limit.

Further, this result fixes some ambiguities.  In classical general relativity, it is possible to use the Friedman equation to replace (for example) $H$ by $\sqrt{8 \pi G \rho / 3}$ in the denominator of $z$; however, the relation between $H$ and $\rho$ changes in LQC.  Therefore, it is necessary to determine what form of $z$ is the correct one for LQC---this calculation based on the separate universe approach gives the preferred form of $z = a \dot\phi/H$ as the correct choice for the potential for scalar perturbations in LQC. (Note that $z$ and $v$ both diverge at the bounce where $H=0$. This is not a problem with the dynamics but rather signals that $v$ is not a good variable for perturbations at the bounce---instead, it is necessary to use another variable, say the comoving curvature perturbation $\mathcal{R}$, to describe scalar perturbations at the bounce point.)

As explained above, the separate universe approach only focuses on super-Hubble modes, and neglects shorter wavelength modes, so a natural question is whether it may be possible to extend this approach to shorter wavelengths. In particular, at the bounce the curvature radius is $\sim \lp$, so is it possible to describe perturbations with a wavelength shorter than $\lp$ by generalizing this approach? (This question is closely related to the trans-Planckian problem that will be discussed in more detail in Sec.~\ref{s.tp}.)

It turns out that this is not possible, for the following reason. In such a quantum theory, for a given state to correspond to a cosmological spacetime with small perturbations it is necessary that the expectation value of any operator $\hat {\cal O}$ evaluated in any cell be close to the expectation value of $\hat{\cal O}$ averaged over all cells. It can be shown that this condition implies that the volume of each cell in the lattice must be much larger than $\lp^3$ \cite{Wilson-Ewing:2012dcf}, so the shortest Fourier mode that can be resolved in this approach for states with small perturbations has a wavelength greater than $\lp$. In short, the approach of discretizing a cosmological spacetime with small perturbations on a lattice works well, but only for perturbations whose wavelengths are always much greater than $\lp$: it is impossible to resolve trans-Planckian modes.

Note that this result does not imply that there must be a Planckian cutoff: perhaps a discretization on a lattice is not an appropriate approximation for trans-Planckian modes, or perhaps it is necessary to allow fluctuations to be large at trans-Planckian scales (although in this case it would be necessary to go beyond linear perturbation theory for these modes). In any case, it does not seem possible to directly generalize the separate universe approach to trans-Planckian modes.

On the other hand, although this has not yet been done, it does seem likely that the separate universe framework could be extended to tensor modes. This could be done by considering a separate universe framework applied to a lattice of Bianchi~I spacetimes (for the loop quantization of the Bianchi~I spacetime, see \cite{Ashtekar:2009vc}), with the metric $g_{ab} = {\rm diag}(-1, a_1(t)^2, a_2(t)^2, a_3(t)^2)$ in each cell of the lattice. Taking $a_3 = a$ and setting $a_1 = a(1 + h)$ and $a_2 = a(1 - h)$ with $h \ll 1$, then $h$ captures a gravitational wave in the $+$ polarization moving in the $x_3$ direction on a flat FLRW background. Using the same steps as for scalar perturbations, it should be possible to derive the equations of motion for this particular tensor mode perturbation. But this result can immediately be generalized, since the same equations should hold for both polarizations of the tensor modes, and for all directions.

In summary, the separate universe approach to cosmological perturbations can be successfully adapted to LQC, giving a loop quantization of long-wavelength scalar modes. This framework has an important strength, as well as a major limitation. Its strength is that this is a loop quantization for these cosmological perturbation modes, not a Fock quantization as is done in the hybrid and dressed metric approaches. On the other hand, the limitation is significant, since it can only be applied to long-wavelength (super-Hubble) modes. As a result, it cannot be used to study inflation, since for inflationary models the observationally relevant modes start with a wavelength much smaller than the Hubble radius. Nonetheless, as shall be discussed later, this approach can be used to study some alternatives to inflation like ekpyrosis and the matter bounce.

\subsection{Anomaly-free effective dynamics}

In the literature, the anomaly-free effective dynamics approach is also referred to as the deformed (or closed) algebra approach. It is a semi-classical approach, in which both homogeneous and inhomogeneous parts of the universe are described by an effective metric with the Lorentz signature, very much like what have been done in classical modern cosmology, with the only exception that quantum corrections from both gravitational and matter sectors (to their leading order) are taken into account. These corrections are obtained by adding quantum counterterms into the classical Hamiltonian constraint, so that the deformed constraint algebra is still closed and anomaly-free. The latter is essential and guarantees that such obtained effective theory is still generally covariant \cite{Teitelboim1973HowCO}.

The anomaly-free effective approach is motivated by results in homogeneous LQC showing that for sharply-peaked states, there exists an effective line element that provides an excellent approximation to the full quantum state \cite{Ashtekar:2006wn, Taveras:2008ke, Diener:2014mia}. Although it is not known if a similar effective description exists for perturbations, general arguments suggest that such a description should be possible for long-wavelength perturbations (but not perturbations with a wavelength $\lesssim \lp$) \cite{Rovelli:2013zaa, Bojowald:2015fla}.

In this approach no wavefunctions are involved, and the usual techniques of classical cosmology can be applied here. In particular, it is possible to work in a general gauge (while the dressed metric and hybrid approaches pick a classically gauge-invariant variable before quantization, and the separate universe approach works in the longitudinal gauge). On the other hand, it is not clear to what extent the inclusion of quantum counterterms in the constraints is unique or not. Finally, although it is in principle possible to include quantum fluctuations in the background spacetime perturbatively \cite{Bojowald:2012xy}, in practice quantum fluctuations are often ignored in this approach.

To describe the deformed algebra approach in more detail, consider the classical constraint algebra, 
\bea
\label{1.15a}
&& \left\{D\left[M^a\right], D\left[N^a\right]\right\} = D\left[M^b\partial_bN^a - N^b\partial_bM^a\right],\\
\label{1.15b}
&& \left\{D\left[M^a\right], S\left[N\right]\right\} = S\left[M^b\partial_bN - N\partial_bM^b\right], \\
\label{1.15c}
&&  \left\{S\left[M\right], S\left[N\right]\right\} = D\left[q^{ab}\left(M\partial_bN - N\partial_bM\right)\right], 
\ea
where $N$ and $M$ are lapse functions, $N^a$ and $M^a$  shift vectors, $q_{ab}$ denotes the spatial 3-dimensional metric, and $D$ and $S$ are the smeared diffeomorphism and Hamiltonian constraints. The constraint algebra is closed and thereby ensures covariance after the (3+1)-dimensional decomposition \cite{Teitelboim1973HowCO}.

When quantum gravitational effects are taken into account, it is expected that this constraint algebra may be modified. Without the full underlying quantum theory, it may be difficult to guess what these modifications may be  \cite{Cuttell:2019fgl}, but to remain covariant the modified algebra should be free of anomalies (that is, the constraint algebra must remain closed) \cite{Teitelboim1973HowCO}.

With this observation in mind, it was found that for linear perturbations on the flat FLRW background, the freedom in the choice of possible deformations can considerably restricted. This was first studied for inverse triad corrections, where under the conditions that (i) the modified effective Hamiltonian constraint must commute with the unchanged (classical) diffeomorphism constraint; (ii) the quantum-corrected  constraints  must form an anomaly-free Poisson algebra,  and (iii) the classical constraints  should be recovered in the classical limit, it was shown that it is possible to find anomaly-free effective constraints for scalar \cite{Bojowald:2008gz, Bojowald:2008jv}, vector \cite{Bojowald:2007hv} and tensor perturbations \cite{Bojowald:2007cd} in closed forms, and the effective equations of motion are derived from these. The observational consequences of these models have been studied, with the result that they can be made consistent with CMB data by properly choosing the parameters of the models \cite{Bojowald:2010me, Bojowald:2011hd, Bojowald:2011iq, Zhu:2015xsa, Zhu:2015owa, Zhu:2015ata}.

Holonomy corrections were considered next, for scalar, vector and tensor modes \cite{Bojowald:2007hv, Bojowald:2007cd, Grain:2009eg, Li:2011zzd, Mielczarek:2011ph, Wilson-Ewing:2011gnq, Cailleteau:2011kr, Cailleteau:2012fy}, with the result that the constraint algebra is modified: while the Poisson brackets \eqref{1.15a}--\eqref{1.15b} are unchanged, the relation \eqref{1.15c} becomes
\bea
\label{1.15d}
&&  \left\{S\left[M\right], S\left[N\right]\right\} = \Omega D\left[q^{ab}\left(M\partial_bN - N\partial_bM\right)\right], 
\ea
where $\Omega \equiv 1 - 2\rho/\rho_c$ (of course, here the smeared Hamiltonian constraint $S[N]$ is the effective version containing holonomy corrections). The observation that $\Omega < 0$ for $ \rho_c/2 < \rho \le \rho_c$ lead to the suggestion of a possible change of signature in the deep ultraviolet  regime \cite{Bojowald:2011aa, Bojowald:2012ux, Bojowald:2015gra} and a possible connection to the no-boundary proposal \cite{Bojowald:2018gdt, Bojowald:2020kob}, although see also a discussion on the claims of signature change in \cite{Wilson-Ewing:2016yan}.

Once the quantum-corrected effective constraints are known, then the equations of motion are derived following the same steps as in general relativity. The anomaly-freedom approach gives a construction for effective constraints that include three classes of terms: zeroth order, first order and second order in the perturbations. The zeroth terms determine the background dynamics with effective Friedman and Raychaudhuri equations, while the first-order terms define the gauge-invariant variables, and the second-order terms generate the dynamics for the perturbations. The equations of motion for the holonomy-corrected scalar perturbations are \cite{Cailleteau:2011kr},
\bea \label{eq25}
v_{k}''(\eta) + \left(\Omega\ k^2 - \frac{z''}{z}\right)  v_{k}(\eta) = 0. 
\ea
For the anomaly-free effective dynamics for tensor modes, see \cite{Cailleteau:2012fy}, and for the inclusion of both holonomy and inverse triad corrections, see \cite{Cailleteau:2013kqa}.

The main drawback of these equations are that they ignore quantum fluctuations so they cannot be trusted to evolve trans-Planckian modes (which necessarily have large quantum fluctuations) \cite{Wilson-Ewing:2016yan, Martineau:2017tdx}. (If one ignores this and imposes initial conditions in the remote contracting phase, for the modes that are trans-Planckian during the bounce, one will obtain power spectra that are  inconsistent with current observations due to significant amplification of the trans-Planckian modes across the bounce when $\Omega < 0$ \cite{Bolliet:2015raa, Grain:2016jlq}.)

Because \eqref{eq25} changes from an elliptic to a hyperbolic equation at $\rho = \rho_c/2$, it has been proposed to fix initial conditions at $\rho = \rho_c/2$ \cite{Mielczarek:2014kea}, this is called the `silent point'. At this point there exists a unique set of initial conditions that give power spectra for the scalar and tensor perturbations that are consistent with the current CMB observations \cite{Li:2018vzr}.

Finally, note that in the long-wavelength limit, the effective scalar perturbation equation  given by \eqref{eq25} is identical to the result derived using the separate universe approach \cite{Wilson-Ewing:2012dcf}, showing the robustness of this choice for the potential $\mathcal{U}$ in LQC.

\section{Predictions for the CMB}
\label{s.app}

The next step in the program is to use the formalisms described above to make connection with the  temperature anisotropies observed in the CMB, as well as to make further predictions testable with current or future observations.  Unfortunately, there is no obvious way in which the bounce of LQC alone can generate the scale-invariant temperature anisotropies of the CMB. Consequently, the strategy so far has been to combine LQC with some other mechanism to generate the primordial perturbations, such as inflation or its alternatives, like ekpyrosis or the matter bounce. In this context, the goal of LQC is not to replace these well-known mechanisms, but rather to complement and extend them to include Planck scale physics. LQC can possibly add some new features to the primordial perturbations which could be used to test this theoretical framework. This section summarizes recent results in this direction.

\subsection{Inflationary models in LQC}
\label{sec:infl}

Among the assumptions on which the inflationary scenario rests, the choice of the initial state for perturbations at the beginning of inflation  is particularly important. The inflationary predictions arise by choosing the so-called Bunch-Davies vacuum at the onset of inflation for the range of wavelengths $\lambda=2\pi/k$ observed in the CMB.  These wavelengths are much shorter than the (spacetime) curvature radius at the onset of inflation ($r_{\rm curv}\approx 1/H$), and the  Bunch-Davies vacuum corresponds to Minkowski-like vacuum fluctuation for these short wavelength modes, plus sub-leading corrections. Although this premise may sound natural at first, it assumes that these modes have never been excited in the past, before inflation starts. This is a strong assumption given our ignorance  about the way  inflation starts and what came before.  Perturbations would not necessarily reach inflation in the Bunch-Davies vacuum if, for instance, the inflationary phase were preceded by a cosmic bounce: then, observable modes could have  exited and re-entered the ``horizon''---the curvature radius, to be more precise---and have been excited during the process. Given a scenario for the pre-inflationary universe, it would be more satisfactory to start the evolution far in the asymptotic past and compute, rather than postulate, the state of perturbations at the onset of slow-roll (although this strategy still requires postulating the initial state for perturbations far in the past). Deviations from Bunch-Davies would carry information about the pre-inflationary evolution, opening an exciting window to explore such a remote era by looking at the CMB. 

This strategy was proposed and explored in LQC in \cite{Agullo:2012sh, Agullo:2013ai}, and further analyzed from different perspectives \cite{Agullo:2015tca, deBlas:2016puz, CastelloGomar:2017kbo, Wu:2018sbr, Zhu:2016dkn, Zhu:2017jew}. We start by listing the main steps in this program, emphasizing the choices and ambiguities at each step. These are:
(1) choice of an inflationary potential $V(\phi)$,
(2) choice of an initial state for the background FLRW quantum spacetime $\Psi_{\rm hom}$ and for
the perturbations $\Psi_{\rm pert}$,
(3) evolution of the perturbations with one of the formalisms described above, and the subsequent computation of observables of interest. We describe now these steps in some more detail.

\noindent
{\it 1. Choice of the inflationary potential.}

At present, there is no compelling candidate for $V(\phi)$  within LQC. This is not surprising; one expects $V(\phi)$ to originate in the matter sector, which is introduced by hand in LQC rather than derived---although one cannot disregard the possibility that the inflaton field and its potential could have a purely gravitational origin \cite{Barrow:1988xi, Barrow:1988xh}. 

The strategy in LQC so far has been the same as in standard inflation: consider  phenomenologically viable potentials and compare their results with observations. Several different forms of  $V(\phi)$ have been analyzed in detail, including the simplest quadratic potential \cite{Agullo:2013ai, Agullo:2015tca}, the Starobinsky potential \cite{Bonga:2015kaa, Bonga:2015xna, Zhu:2017jew}, monodrony \cite{Sharma:2018vnv} and $\alpha$-attractor potentials \cite{Shahalam:2019mpw}. 
Of course, it is important to distinguish genuine LQC effects from those features arising from a concrete choice of $V(\phi)$.

As first discussed in \cite{Ashtekar:2009mm, Ashtekar:2011rm}, and further analyzed in \cite{Corichi:2010zp, Martineau:2017sti, Li:2018fco}, in presence of a viable inflationary potential $V(\phi)$ the dynamics of the scalar field across the bounce of LQC  can set up the appropriate conditions for inflation to start, quite generally if the kinetic energy of the scalar field dominates over its potential energy at the bounce. 
In this concrete sense, the attractor character of inflation in general relativity persists in LQC, if one assumes an appropriate potential for $\phi$.

\noindent
{\it 2. Choice of the initial quantum state.}

To ensure a good semi-classical limit, the quantum background FLRW spacetime $\Psi_{\rm hom}(a,\phi)$ is typically chosen to be a sharply-peaked state, whose main features are captured by the effective equations of LQC. In this case, the freedom in the choice of solution is quite simple, and in fact reduces to a one parameter freedom that dictates the length of the inflationary phase (as measured in number of $e$-folds $N$). This point will be important when discussing predictions for the CMB.

Computing predictions for the CMB also  requires choosing the state of perturbations, $\Psi_{\rm pert}$, at some initial time, which can then be evolved across the pre-inflationary and inflationary phases. The predictions for the primordial power spectrum crucially depend on this choice. In the absence of a compelling way to specify the initial state, the theory would lack predictive power. The vacuum state is a natural choice, but, as is well known from quantum field theory in time-dependent spacetimes, the notion of vacuum is ambiguous except in very special circumstances. The strategy in LQC has been to add physical arguments to single out a choice, and then compare the resulting predictions with observations; this may provide some confidence with the choice made, or could rule it out. Thus, observations test not only the theory of LQC, but also the arguments on how to  fix the initial sate of perturbations. 

Multiple strategies have been proposed and explored in this regard. Perhaps the most conservative strategy is to fix the initial state of perturbations in the far past before the bounce. In addition of being a natural strategy in a bouncing universe, it has the advantage that far in the past and under mild assumptions, perturbations are far inside the curvature radius and therefore all different notions of natural adiabatic vacua converge (for the same reason that everyone agrees on the vacuum state in labs on Earth, even though the universe is expanding) \cite{Bolliet:2015bka, Schander:2015eja, Bolliet:2015raa, Agullo:2015tca, Agullo:2017eyh, Olmedo:2018ohq, Agullo:2021oqk, Agullo:2020wur, Agullo:2020iqv, Li:2020mfi,Zhu:2017jew}. Notice  that  the same strategy is also used in alternatives to inflation, such as ekpyrosis and the matter bounce discussed in  Secs.~\ref{ekpyrosis} and \ref{matterbounce}, since in these scenarios the primordial power spectrum is generated before the bounce and therefore one must specify the initial state far in the contracting branch. 

Another possibility is to use the bounce as a preferred time to specify the initial conditions \cite{Agullo:2012sh, Agullo:2012fc, Agullo:2015tca, CastelloGomar:2017kbo, Martin-Benito:2021szh, Zhu:2017jew}. In this strategy, there is some ambiguity on the choice of initial state for a range of the smallest wavenumbers (longest wavelengths) we can observe in the CMB, since in scenarios of phenomenological interest these wavelengths are greater than the curvature radius at the bounce. Different proposals for a preferred vacuum state at (or near) the bounce have been considered so far using two arguments, namely the extrapolation of the adiabatic series \cite{Agullo:2012sh, Agullo:2013ai, Agullo:2015tca} and minimization conditions for the expectation value of the energy-momentum tensor \cite{Agullo:2014ica, Martin-Benito:2021szh}. Interestingly, although    prescriptions using either of these two strategies differ significantly in their details, they all produce very similar observational predictions \cite{Agullo:2015tca, Martin-Benito:2021szh}, which are also quite similar to the results obtained when specifying adiabatic initial conditions far in the past of the bounce. 

A third strategy that has been proposed to fix the state of perturbations is to impose conditions on $\Psi_{\rm pert}$ at more than one instant, so these conditions are non-local in time---e.g., conditions  the state must satisfy both at the bounce and the end of inflation \cite{Ashtekar:2016pqn, Ashtekar:2016wpi, ElizagaNavascues:2020fai}, or during an interval of the pre-inflationary evolution \cite{deBlas:2016puz, Martin-Benito:2021ulw}.  The two existing proposals based on this third strategy do predict a primordial power spectra substantially different from the ones obtained with either of the other two strategies summarized above; we discuss this further at the end of Sec.~\ref{s.inf-app}.

\noindent
{\it 3. Computation of predictions in the different frameworks.}

In the remainder of this section, we provide a summary of the results obtained using the different frameworks described in Sec.~\ref{s.perts-lqc}, describing the predictions for observables of interest in primordial cosmology, including the amplitude of scalar and tensor primordial perturbations, their spectral indices and runnings. Non-Gaussianity and primordial anisotropies are discussed in Sec.~\ref{s.ext}. Particular attention has also been paid to the so-called large-scale anomalies in the CMB \cite{Planck:2019evm}.

\subsection{Inflation in the dressed metric and hybrid approaches}
\label{s.inf-app}

Since the dressed metric and hybrid quantization approaches give very similar predictions for the primordial scalar and tensor power spectra in inflation, we discuss them together here; see \cite{Li:2022evi} for a detailed comparison between the two approaches.

Before describing the results for the power spectrum, it is informative to acquire first an intuitive understanding of the type of modifications LQC can cause, relative to standard inflation, and their physical origin. 

The argument is simplest for tensor modes $\chi_k$ (with $\chi$ the analog for tensor perturbations of the Mukhanov-Sasaki variable $v$) since ${\cal U}_{\rm tensor} = 0$; the equation of motion is slightly more complicated for scalar perturbations, but the general argument remains the same. Since ${\cal U}_{\rm tensor} = 0$, and recalling that the Ricci curvature of the dressed metric is $R=6 {\tilde a''}/{\tilde a^3}$, the equation of motion for tensor modes is
\be \label{chieqn}
\chi''_k(\eta)+ \tilde a^2 \, \left(\frac{k^2}{{\tilde a}^2} - \frac{1}{6}\,  R\right)\, \chi_k(\eta)=0\, .
\ee
This equation shows that the evolution of $\chi_k$ is dictated by a competition between the physical wavenumber squared of the Fourier mode $k$ and the Ricci scalar curvature. If $(k / \tilde a)^2 \gg R$, $R$ can be neglected and then $\chi''_k(\eta)+k^2\, \chi_k(\eta)=0$, which is the equation we would have found in Minkowski spacetime and whose solutions are linear combinations of positive and negative frequency modes $e^{\pm i \, k \, \eta}$. It is known from quantum field theory in Minkowski spacetime that this simple evolution does  not create particles or excite the vacuum state. Restated in terms of wavelengths, the Fourier modes whose wavelength are much smaller that the ``radius of curvature'' $r_{\rm curv}=\sqrt{6/R}$, behave as in flat spacetime, and vacuum fluctuations remain unexcited.

On the contrary, when the physical wavenumber squared becomes comparable to the curvature $(k / \tilde a)^2 \sim R$, the effective frequency of oscillation of $\chi_k$ becomes time-dependent and complex exponential  $e^{\pm i \, k \, \eta}$ are no longer solutions; this is the regime where perturbations are affected by the curvature of the background spacetime.

\begin{figure}[tb] \begin{center}
  \includegraphics[width=4in]{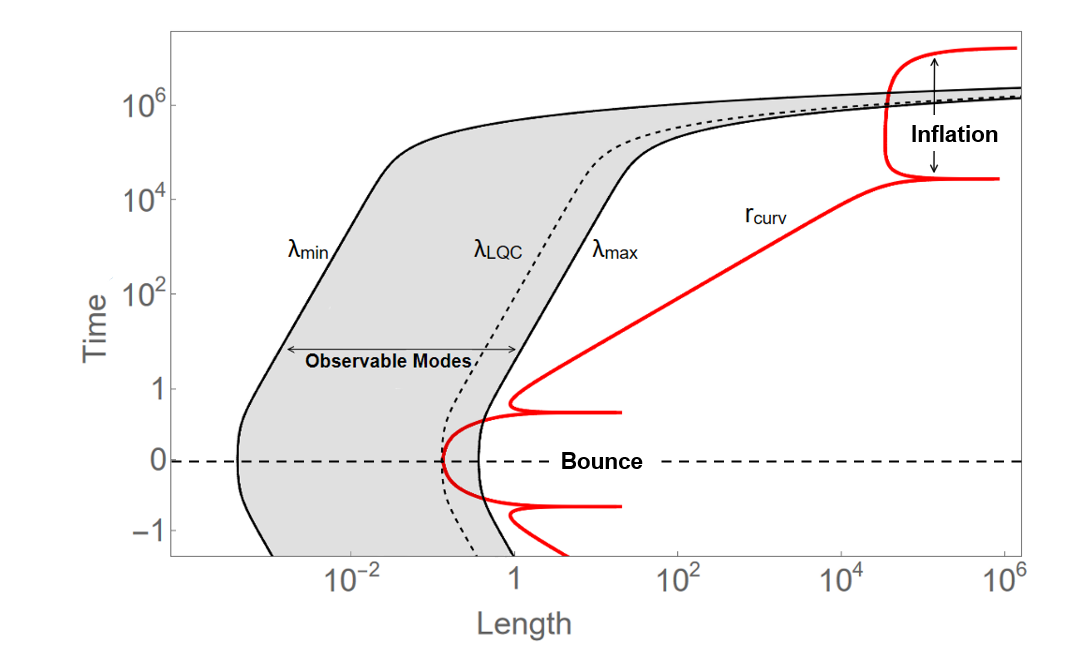}  \end{center}
    \caption{This plot shows the LQC bounce followed by a phase of inflation. The ``curvature scale'' $r_{\rm curv}(t)$ is shown as a red line, while the gray region corresponds to the range of wavelengths observed in the CMB today. (Note that $r_{\rm curv}(t)$ diverges when the Ricci scalar changes sign; this happens just before and after the bounce, as well as before and after inflation.)
    In this scenario, the longest wavelength modes observed in the CMB today were comparable to or larger than the curvature scale near the bounce. Consequently, these modes were excited at that time, and can provide an observational window to the Planck era.} \label{fig:diagram}
\end{figure}

To get an idea of when primordial perturbations may become excited, it is sufficient to compare the evolution of their physical wavelength $\lambda(t)$ with the curvature radius $r_{\rm curv}(t)$. An example of this is shown in Fig.~\ref{fig:diagram}, for the case of a quadratic potential  $V(\phi) = \frac{1}{2} m^2 \phi^2$ and a choice of initial conditions that produces a total of 68 $e$-folds of inflation. The red line shows the radius of curvature from some time before the bounce until the end of inflation, while the gray shadowed band indicates the range of wavelengths observed in the CMB. For this concrete evolution, the longest wavelengths observed today in the CMB become larger than the curvature radius near the time of the bounce---during this interval, perturbations with these wavelengths are affected by the background spacetime curvature and as a result, when inflation starts after the bounce these wavelengths are already in an excited state compared to the Bunch-Davies vacuum. On the other hand, the shortest wavelengths observed in the CMB today remain much smaller that $r_{\rm curv}$ during the entire Planck era, and only become comparable to $r_{\rm curv}$ at much later times during inflation, so these wavelengths will reach the onset of inflation in the vacuum state.

The division between wavelengths that ``feel'' the geometry during the bounce, and those that do not, is determined by the value of the curvature radius at the bounce $r_{\rm curv}(t_B)$: this is the physical scale that LQC introduces in the physics of primordial perturbations. Wavelengths satisfying $\lambda(t_B) \gtrsim r_{\rm curv}(t_B)$ will carry some information about LQC. In terms of  comoving wavenumbers, the modes with $k\lesssim k_{\rm LQC}$, where $k_{\rm LQC} \equiv 1 / r_{\rm curv}(t_B)$, are the ``messengers" from the Planck era.

As mentioned above, this discussion can be extended to scalar modes with essentially the same result, as the potential $\tilde{\mathcal{U}}$ does not substantially change the argument. 

Leaving aside qualitative arguments, it is possible to solve the dynamics explicitly numerically (for analytic results, see \cite{Zhu:2017jew, Li:2018vzr, Zhu:2015ata, Zhu:2015owa, Zhu:2015xsa, Wu:2018sbr}) and calculate the observables of interest, for example the power spectra at the end of inflation.

\begin{figure}[tb] \begin{center}
  \includegraphics[width=4in]{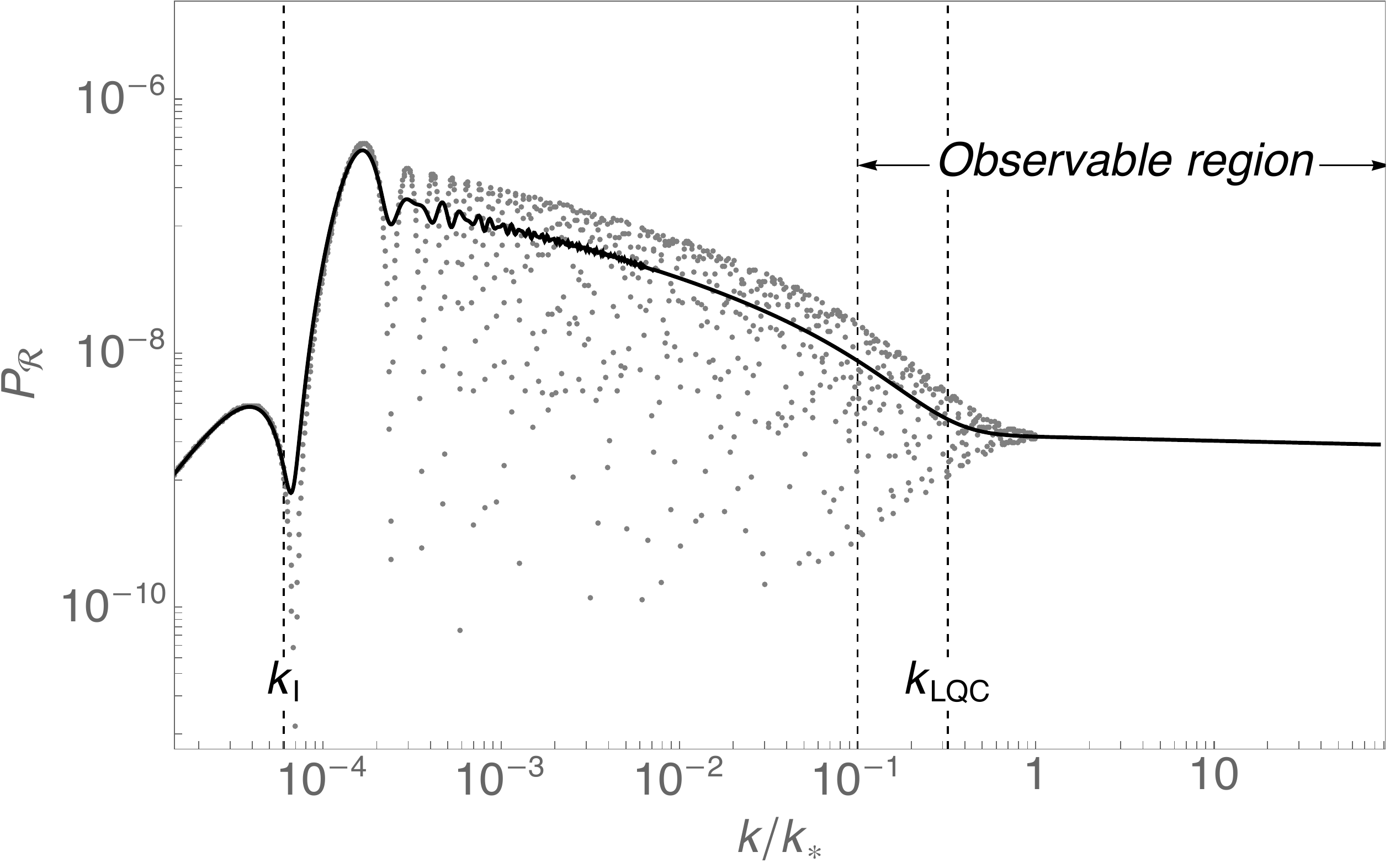}  \end{center}
    \caption{Scalar power spectrum in LQC versus wavenumber $k/k_*$, where $k_*$ is a reference  comoving scale corresponding to $0.002\, {\rm Mpc}^{-1}$ today. The LQC characteristic scale $k_{\rm LQC}$ is denoted by a vertical line, and the gray dots show the scalar power spectrum, computed for a discrete set of wavenumbers. The power spectrum oscillates rapidly and its average is shown in black. It is almost scale-invariant and in agreement with the predictions of standard  inflationary with Bunch-Davies initial conditions for $k>  k_{\rm LQC}$. On the contrary, for $k<  k_{\rm LQC}$ the near scale-invariance is broken due to LQC effects that excited these modes during the Planck era. The plot also shows, for a certain choice of parameters, the region of Fourier modes that are observable in the CMB as well as $k_I$, the most infrared mode that exits the horizon during the inflationary phase of the universe \cite{Agullo:2015tca}.
    } \label{fig:spectra}
\end{figure}

As an example, Fig.~\ref{fig:spectra} shows the scalar power spectrum calculated for a particular scenario and choice of initial conditions. Specifically, the inflationary potential chosen here is $V(\phi)=\frac{1}{2}\, m^2\, \phi^2$ with $m=1.3 \times 10^{-6}$ has been used in this plot, (note that the qualitative result for LQC effects on the power spectrum is very similar for other inflationary potentials) and initial conditions for the background such that there are about 4 $e$-folds from the bounce to the onset of inflation, and approximately 64 $e$-folds of inflation, while the initial conditions for the perturbations are obtained  using  the so-called preferred instantaneous vacuum \cite{Agullo:2014ica} defined at $t_i = -50000~t_{\rm Pl}$ before the bounce (this is a vacuum state of fourth adiabatic order and the choice of $t_i$ does not affect the results so long as it is far before the bounce). This plot shows that, as expected, LQC effects appear on infrared scales $k <  k_{\rm LQC}$, while predictions for the modes $k > k_{\rm LQC}$ are unchanged from the standard inflationary scenario. In this scenario, modes modified by LQC can be observed in the CMB, although only in the most infrared sector.  Concretely, LQC modifications appear for $\ell \lesssim 30$ in the angular power spectrum $C_{\ell}$ for this choice of parameters. Results for tensor modes are similar, although with a smaller amplitude \cite{Agullo:2014ica}.

In summary, LQC introduces a new physical scale $k_{\rm LQC}$ in the physics of primordial perturbations. The scale invariance of the power sepectrum generated by inflation is modified for modes $k < k_{\rm LQC}$: the spectral indices of both scalar and tensor perturbations become more negative or, equivalently, the running of the two spectral indices increases. Note that the modifications that LQC introduces for scalar and tensor modes are very similar, so the tensor-to-scalar ratio $r$ remains the same as in standard inflation. This implies that the consistency relation $r=-8\, n_t$, where $n_t$ is the tensor spectral index, valid in standard inflation, is not satisfied in LQC for modes $k < k_{\rm LQC}$ for which $r < -8\, n_t$ instead \cite{Agullo:2013ai, Agullo:2015tca}. 

Similar results have been obtained using other choices for the quantum state of perturbations, specified at or before the bounce \cite{Agullo:2015tca}, although some quantities like the slope of the power spectrum in the intermediate region change slightly \cite{Li:2019qzr}. In addition, choices motivated by arguments that are non-local in time \cite{Ashtekar:2016pqn, ElizagaNavascues:2020fai, deBlas:2016puz, Martin-Benito:2021ulw} produce LQC corrections that reduce the primordial power spectrum, rather than enhance it. The sign of the spectral indices and runnings is, therefore, reversed if one used one of theses states based on the non-local conditions, and the consistency relation is modified in the inverse manner to $r> -8\, n_t$ for infrared wavenumbers \cite{Ashtekar:2016pqn, deBlas:2016puz, Martin-Benito:2021ulw}. \\

\newpage

\noindent
{\bf Large-scale anomalies in the CMB} \\

Since LQC modifies the primordial power spectrum at infrared scales, this raises the question of whether LQC can provide a natural mechanism to account for the anomalous features  observed in the correlation of temperature anisotropies at large angular separations; these are called large-scale  anomalies in the CMB (see, e.g., \cite{Schwarz:2015cma} for a summary).
These anomalies include: (i) the absence of two-point correlations at large scales, also known as power suppression; (ii) a hemispherical or dipolar asymmetry;  (iii) a bias for odd-parity correlations; and (iv) a preference of data for a value of the lensing amplitude $A_L$ larger than one \cite{Akrami:2018vks}. Most of these signals have been detected in data from the satellites WMAP and Planck, and some have been noticed even in data from COBE; this rules out an origin from instrumental noise or residual systematics. There is  agreement that these signals are real features in the CMB. The discussion is instead whether the observed signals require new physics. The significance of these features has been quantified by using the $p$-value \cite{Ade:2013nlj}, that measures the probability of observing each of these features in a standard $\Lambda$CDM universe. The three anomalies separately have similar $p$-values,  of the order of a fraction of per cent \cite{Ade:2015hxq,Planck:2019evm}. These $p$-values are small, but not sufficiently small to close the discussion about the need of new physics.  It is important to notice though that these are the $p$-values of each of the anomalies separately. Their collective $p$-value must be smaller, so their collective significance is higher. 

Within LQC, there have been two different proposals to account for these anomalies, which we summarize in the the following. 

The first proposal points out that the non-Gaussianty generated by the LQC bounce can make the observed features much more likely than in standard $\Lambda$CDM \cite{Agullo:2015aba, Agullo:2021oqk}. The underlying idea is based on the ``non-Gaussian modulation'' mechanism \cite{Schmidt:2010gw, Schmidt:2012ky, Jeong:2012df, Dai:2013kfa, Agullo:2015aba, Adhikari:2015yya} where correlations between CMB modes and super-horizon modes can bias the observed power spectrum, making certain features to be more likely. An important challenge for non-Gaussian modulation is to generate an appropriate form of non-Gaussianity that is small when the three wavenumbers lie within the observable window, in order to respect observational constraints, but that grows significantly when one of the modes is super-horizon. It is interesting that these are precisely the properties of the non-Gaussianity generated by the LQC bounce \cite{Agullo:2017eyh} (for details, see Sec.~\ref{nonG}), in such a way that the resulting non-Gaussian modulation due to LQC can alleviate the four large-scale anomalies mentioned above \cite{Agullo:2021oqk, Agullo:2020fbw, Agullo:2020cvg}. The way this alleviation happens is statistical: LQC does not predict that such anomalies must be necessarily present in our universe, but rather makes their $p$-values significantly larger than in $\Lambda$CDM, so the reason why these features were called anomalies in the first place disappears. (The analysis also makes predictions for tensor modes which can be contrasted with observations if these modes are eventually measured.)

The main strength of this idea is that it can account for several anomalies, of very different nature, like the power suppression and the dipolar asymmetry, while the main limitation is its inability to incorporate the rapid oscillations of the LQC bispectrum. Since these oscillation may reduce the effects of non-Gaussianity in the observed CMB, the results of \cite{Agullo:2021oqk} must be understood as an upper bound for the significance of the anomalies within LQC, rather than a sharp prediction. We comment on other possible observational signatures of this scenario for non-Gaussianities in Sec.~\ref{nonG}.

The other avenue explored within LQC to account for some of the anomalies is based on `initial' conditions for the perturbations that are non-local in time. Specifically, for the quantum state singled out by demanding that the power spectrum at the end of inflation does not  oscillate in $k$, the scalar power spectrum is suppressed in the infrared part of the visible window \cite{deBlas:2016puz}. A similar result was obtained by using a quantum state for scalar perturbations obtained by combining a quantum generalization of Penrose's null Weyl curvature hypothesis with the demand that the state has minimum uncertainty in the curvature perturbations (and therefore maximum uncertainty in the canonically conjugated momentum) at the end of inflation \cite{Ashtekar:2016pqn, Ashtekar:2016wpi}. Further, the authors of \cite{Ashtekar:2020gec,Ashtekar:2021izi} noticed that this suppression can also alleviate the observed anomaly in the lensing parameter $A_L$, by making the observed value compatible with 1 within 1$\sigma$; this analysis comes with new predictions of a larger optical depth and power suppression for the $B$-mode polarization.

\subsection{A first look at quantum ambiguities: inflation in modified LQC}
\label{s.modlqc}

An outstanding issue in LQC is the connection between it and the full theory of LQG. The starting point for LQC is to reduce the classical Hamiltonian from infinitely many to a few gravitational degrees of freedom  by imposing  homogeneity (isotropy further reduces the system to a single degree of freedom) and the reduced system is quantized using the techniques of LQG. However, the processes of symmetry reduction and quantization do not commute in general, and it is important to understand how well the physics of the cosmological sector of full LQG is captured by LQC. In the past decade or so, this important issue has been extensively studied by both bottom-up and top-down approaches; from these studies, an important conclusion is emerging: {\em LQC and its major predictions are robust}. In particular, the big bang singularity is resolved in all the models studied so far, and predictions for cosmological perturbations are consistent with current cosmological observations for a wide variety of initial conditions.

We will review results from the top-down approach in Sec.~\ref{s.lqg}, while we focus here on the bottom-up approach. In this setting, symmetries are still imposed before quantization, but with the observation that the Hamiltonian constraint can be written in different equivalent forms at the classical level. In one commonly used form, the constraint contains two terms often called `Euclidean' and `Lorentzian' respectively (since only the first term appears in the Hamiltonian constraint for Euclidean gravity). For the spatially flat FLRW universe, these two parts are proportional in the classical theory, and this simplification is typically used in LQC to reduce the total Hamiltonian constraint to a single Euclidean term (although with a different factor of proportionality). Since the Euclidean and Lorentzian terms are usually regularized differently in LQG, this motivates treating the Lorentzian term independently by applying
Thiemann's regularization \cite{Thiemann:1996aw} of the full theory of LQG to LQC \cite{Yang:2009fp},
with the result that the wave-function is described by a fourth-order difference equation \cite{Yang:2009fp,Dapor:2017rwv}, rather than the second-order  difference equation appearing in standard LQC. For sharply peaked states, the resulting quantum dynamics are well described by effective Friedman-Raychaudhuri equations \cite{Li:2018opr, Li:2018fco, Li:2019ipm}.

This version of LQC is often called mLQC-I, with `m' for modified \cite{Li:2018opr}. Note that another modified version of LQC, called mLQC-II, is obtained by imposing that the spin-connection vanishes before quantizing the Euclidean and Lorentzian terms separately \cite{Yang:2009fp}. It has since been shown that the physics of mLQC-II is very similar to standard LQC \cite{Li:2018opr, Li:2018fco, Li:2019ipm, Li:2020mfi, Li:2019qzr, Li:2021mop}, and therefore we will focus on mLQC-I here.

Comparing mLQC-I with LQC, one finds that the big bang singularity is still replaced by a quantum bounce, but the critical energy density the bounce occurs at in mLQC-I is smaller by a factor of $4(1+\gamma^2)$. More significant differences arise comparing the pre-bounce eras: in standard LQC the classical Einstein dynamics are recovered far from the bounce, both to the past and future, while in mLQC-I this is only true for one side of the bounce. If we assume the post-bounce era to have a good classical limit to match our expanding universe, then in mLQC-I the pre-bounce universe rapidly approaches a de Sitter phase, with a Planckian effective cosmological constant. Despite the pre-bounce differences, the post-bounce dynamics of standard LQC and mLQC-I are essentially identical \cite{Li:2018opr, Li:2018fco, Li:2019ipm}.

In mLQC-I, an inflationary phase generally occurs, assuming an inflaton with a suitable potential \cite{Li:2019ipm}. If the inflaton is kinetic-dominated at the bounce, $\dot{\phi}_B^2/2 \gg V(\phi_B)$, then inflation always happens at $t_{i} \simeq 10^{4} - 10^{6}~t_{\rm Pl}$. On the other hand, if the scalar field is initially dominated by its potential energy at the bounce, inflation does not always happen; this is true not only in mLQC-I \cite{Li:2018opr, Li:2018fco, Li:2019ipm} but also in standard LQC \cite{Bonga:2015kaa,Bonga:2015xna}. Note that kinetic-dominated initial states can be expected to arise quite generally, given the Hubble anti-friction term in the Klein-Gordon equation for a contracting universe which will significantly increase $\ddot\phi$.

The evolution of the homogeneous universe is simple and universal for initial states dominated by kinetic energy at the LQC bounce. In terms of the effective equation of state of the inflaton, $w(\phi) \equiv [\dot{\phi}^2 - 2V(\phi)]/ [\dot{\phi}^2 + 2V(\phi)]$, it has the value $w(\phi) \simeq 1$
during a long kinetic-dominated era (of $\Delta t \simeq 10^{5} t_{\rm Pl}$), and the potential energy remains nearly constant. When $t - t_B \simeq 10^{5} t_{\rm Pl}$, the kinetic energy suddenly drops and $w(\phi) \rightarrow -1$, signaling the start of inflation when the potential energy takes over. Therefore, there are three phases before reheating: the bounce, the kinetic transition and then inflation \cite{Li:2018opr,Li:2018fco,Li:2019ipm}. Note that similar behavior has been also found in standard LQC for kinetic-dominated initial conditions for the inflaton at the bounce \cite{Zhu:2016dkn, Zhu:2017jew}, and is true for a wide range of inflationary potentials \cite{Shahalam:2017wba, Shahalam:2018rby, Sharma:2018vnv, Bhardwaj:2018omt, Shahalam:2019mpw, Sharma:2019okc}.

For a systematic study of mLQC-I, see \cite{Assanioussi:2018hee, Assanioussi:2019iye}. In addition, also within the minisuperspace approach of LQC, a reduced phase space quantization with an inflaton field and several different reference fields recovers many results of LQC, including showing that the resolution of the big-bang singularity is robust \cite{Giesel:2020raf}.

Cosmological perturbations have been studied in mLQC-I \cite{Agullo:2018wbf, Garcia-Quismondo:2020wna, Gomar:2020orw, Li:2020mfi, Li:2019qzr, Li:2021mop}, with the equations for the scalar and tensor perturbations the same as for standard LQC, with the only difference that in mLQC-I the effective background geometry (particularly the pre-bounce era) is different \cite{Li:2018opr, Li:2018fco, Li:2019ipm}. For the contracting phase, the background is well approximated by de Sitter contraction with $ \left|aH\right| = -\eta^{-1}$, and the equations of the scalar and tensor perturbations reduce to those given in general relativity. 
Using the same arguments as in semi-classical cosmology, a natural choice for the initial state of perturbations in the contracting de Sitter space is the Bunch-Davies vacuum \cite{BD78}
\be \label{eq34}
v_k^{\text{(initial)}} =  \frac{1}{\sqrt{2k}}e^{-ik\eta}\left(1 - \frac{i}{k\eta}\right). 
\ee
In classical slow-roll inflation, the background is also almost de Sitter, and at sufficient early times $a(\eta) \simeq 1/(-H\eta) \ll 1$ so $|\eta k| \simeq |H\eta| \gg 1$ and therefore the term ${i}/{k\eta}$ in \eqref{eq34} is negligible; as a result the modes \eqref{eq34} become indistinguishable from  those defining the Minkowski vacuum $v_k^{\text{(Mink.)}} = e^{-ik\eta}/{\sqrt{2k}}$. In contrast, for mLQC-I, far in the past contracting phase the modes of interest lie outside the Hubble radius, so the term $i/k\eta$ in \eqref{eq34} cannot be neglected; for more details, see \cite{Li:2021mop}.

The ambiguity in the form of the effective potential $\mathcal{U}$ appearing in the equation of motion of scalar perturbations (see the discussion below Eq.~\eqref{hybrid1a} on the origin of this ambiguity) remains in  mLQC-I. However, contrary to standard  LQC where different choices produce quite similar results for the power spectrum \cite{Agullo:2013ai, Agullo:2015tca}, this is not the case for mLQC-I \cite{Li:2020mfi, Li:2019qzr}; in fact some choices have been already ruled out by current observations \cite{Li:2019qzr}.

Imposing Bunch-Davies initial conditions in the remote contracting phase and using the dressed metric approach, it was found that (similarly to standard LQC) the power spectra of the cosmological scalar and tensor perturbations can be divided into three regimes, ultraviolet, oscillatory and infrared, respectively corresponding to wavenumbers $k > k_{\rm LQC}$, $k_{\rm I} < k < k_{\rm LQC}$, and $k < k_{\rm I}$ (see Fig.~\ref{fig:spectra} for the definition of $k_{\rm LQC}$ and $k_I$) \cite{Agullo:2018wbf, Li:2019qzr, Garcia-Quismondo:2020wna}. The major difference between the power spectra obtained in mLQC-I and standard LQC lies in the oscillatory and infrared regimes.
First, in the oscillatory regime the power spectrum of the scalar and tensor perturbations is proportional to $k^{-3}$ in mLQC-I, as compared to $k^{-1}$ in  LQC \cite{Agullo:2018wbf}; however, this property does depend on the initial conditions of the scalar field and the choice of the potential \cite{Li:2019qzr}.
Second, the scalar power spectrum was also studied in the hybrid approach \cite{Garcia-Quismondo:2020wna, Gomar:2020orw, Li:2020mfi} and, although the three different regimes mentioned above are also present in this case, some differences were found in the infrared and oscillatory regimes of mLQC-I, where a suppressed power spectrum was found  for these infrared modes \cite{Li:2020mfi}. 
This is in striking contrast with the results of the dressed metric approach where the power spectrum is amplified for precisely these modes. Nevertheless, these differences arise only for the modes in the oscillatory and infrared regimes, while for the modes  in the observable window (i.e., the ultraviolet regime) the differences are  less than $1\%$ \cite{Li:2021mop}. Since the modes observed in the CMB today belong  mainly to the ultraviolet regime, it may be difficult to distinguish LQC from mLQC-I observationally, as well as the dressed metric and hybrid quantization approaches that differ in mLQC-I. It is possible that other observables like non-Gaussianities may be able to distinguish these scenarios, this is a topic of current research.

\subsection{Ekpyrosis in LQC}
\label{ekpyrosis}

Ekpyrosis is an alternative to inflation based on postulating  the existence of a period of slow contraction due to the presence of an ultra-stiff fluid with $p > \rho$; for a review see \cite{Lehners:2008vx}. Ekpyrosis is often taken to be a cyclic cosmology, with multiple recollapse and bounce cycles, although to generate perturbations that can match the observations of the CMB only one collapse and bounce phase is necessary.

Assuming there are multiple matter fields, a phase of ekpyrosis will generate nearly scale-invariant entropy perturbations (with a small red tilt), and these can in turn act as a source to excite density perturbations with the same nearly scale-invariant spectrum \cite{Finelli:2002we, DiMarco:2002eb, Lehners:2007ac, Koyama:2007mg} to match observations.  On the other hand, tensor modes are never significantly excited during ekpyrosis \cite{Boyle:2003km}, so a clear prediction for ekpyrotic cosmologies is a vanishingly small tensor-to-scalar ratio.

Although the ekpyrotic scenario was first proposed in string-inspired braneworld cosmology \cite{Khoury:2001wf}, the key ingredient to generate scale-invariance---slow contraction due to an ultra-stiff fluid---is entirely independent of string theory. For example, in LQC it is possible to couple scalar fields with an appropriate potential to obtain the cyclic dynamics typically expected for ekpyrosis \cite{Cailleteau:2009fv}.

Another essential ingredient needed for any ekpyrotic model is a bounce, which can be provided by LQC. Further, since observationally relevant modes are super-horizon during the bounce, the equations of motion for the perturbations through the bounce can be taken from the separate universe quantization for long-wavelength modes \cite{Wilson-Ewing:2012dcf, Wilson-Ewing:2015sfx}.

Solving the dynamics shows that the (nearly) scale-invariant curvature perturbations travel through the bounce unscathed, and they freeze once the background spacetime starts to expand after the bounce \cite{Wilson-Ewing:2013bla}. This demonstrates that LQC can complete the ekpyrotic paradigm, providing the bounce necessary to pass from a phase of slow contraction to our currently expanding universe. On the other hand, LQC does not modify the predictions in any way, so while ekpyrotic LQC is a viable cosmology there are no predictions for LQC-specific effects in the CMB---the predictions for ekpyrosis are independent of the bounce mechanism.

\subsection{LQC Matter Bounce}
\label{matterbounce}

Another alternative to inflation that relies on a cosmic bounce is the matter bounce scenario. In a contracting universe, if the dynamics are dominated by a matter (also often called dust) field with vanishing pressure, then cosmological perturbations become scale-invariant as they exit the Hubble radius, assuming the modes were initially in the vacuum quantum state \cite{Wands:1998yp}. If there is a cosmic bounce, then these scale-invariant perturbations can provide appropriate initial conditions for the CMB \cite{Finelli:2001sr}. A slight red tilt for the scalar power spectrum is obtained if the equation of state $w = p/\rho$ of the matter content is slightly negative \cite{Wilson-Ewing:2012lmx}, for example due to a small contribution from dark energy \cite{Cai:2014jla, Cai:2016hea}.

Since LQC automatically provides a bounce, with a matter-dominated era of contraction it is one possible realization of the matter bounce scenario \cite{Wilson-Ewing:2012lmx}. The simplest version of the LQC matter bounce is to assume a matter-domainted phase all the way to the bounce, but this scenario faces several difficulties. First, in this case the predicted amplitude of the scalar perturbations is determined by the energy density at the bounce (in natural units), so a Planckian bounce gives perturbations of order one and is clearly ruled out by observations (in addition to the fact that the regime of linear perturbation fails in this case). This problem can be avoided by requiring that the bounce occur at scales far below the Planck scale, but then this seems somewhat unnatural from a quantum gravity perspective. Second, a matter-dominated contracting universe is unstable to the growth of anisotropies \cite{Levy:2016xcl}, which also suggests that a matter bounce scenario with vanishing pressure for the entire contracting phase requires extensive fine-tuning in the initial conditions for anisotropies to always remain small.

A scenario that alleviates these problems is to have the era of matter contraction followed by a period of ekpyrotic contraction before the LQC bounce \cite{Cai:2014zga, Li:2020pww}. Then, the amplitude of the scalar perturbations is determined by the energy scale at the transition time between matter and ekpyrotic contraction (not the energy scale of the bounce) \cite{Qiu:2013eoa, Cai:2013kja, Cai:2014zga}, and the energy density of the ekpyrotic fluid grows more rapidly than anisotropies in a contracting universe, also alleviating the anisotropy problem \cite{Cai:2012va, Cai:2013vm}. Even in this case, there are observational constraints from CMB data that provide very strong bounds on the possible strength of primordial anisotropies at the bounce \cite{Agullo:2022klq}, indicating that the anisotropy fine-tuning problem, although alleviated, is not entirely avoided by adding an ekpyrotic phase of contraction.

Another strong observational constraint on the matter bounce scenario are the latest bounds on the tensor-to-scalar ratio of $r < 0.036$ \cite{BICEP:2021xfz}. The simplest realizations of the matter bounce scenario typically predict a value for $r$ close to 1, which is clearly ruled out. There are ways to suppress $r$, for example by including several matter fields, setting the sound speed of the matter field to be small, or amplifying the scalar perturbations during the bounce, but these typically generate non-Gaussianities that violate observational bounds \cite{Cai:2011zx, Quintin:2015rta, Li:2016xjb}.

Interestingly, the LQC bounce can suppress the tensor-to-scalar ratio by decreasing the amplitude of tensor perturbations during the bounce \cite{Wilson-Ewing:2015sfx}, thereby alleviating this problem. The factor of suppression depends on the matter field at the bounce; for example, $r$ decreases by a factor of 4 for radiation-domination, and the suppression is stronger the closer the equation of state $w$ is to 0 during the bounce. Note, however, that to significantly decrease $r$ it is necessary for the equation of state to be small, which reintroduces the anisotropy problem discussed above.

In summary, LQC provides a simple realization of the matter bounce scenario, and LQC has the beneficial effect that it can successfully decrease the predicted tensor-to-scalar ratio. Despite this, it remains a challenge for the matter bounce scenario (whether realized in LQC or in some other bouncing cosmology) to simultaneously satisfy observational constraints from the CMB on the tensor-to-scalar ratio, non-Gaussianities, and  primordial anisotropies.

\section{Extensions}
\label{s.ext}

The discussion so far has focused on the simplest case: linear perturbations on an isotropic background. In this section, we discussion extensions to go beyond linear perturbation theory and allow for an anisotropic background spacetime.

\subsection{Non-Gaussianity}
\label{nonG}

The study of cosmological perturbations so far has been based on linear perturbation theory. Going to the next order in perturbation theory is important to demonstrate the viability of the theoretical framework by verifying that the perturbation theory remains under control, and the compatibility of its predictions with observations. This is a non-trivial demand, since self-interactions between perturbations are mediated by effective couplings of gravitational origin, and these couplings could become large when curvature invariants reach the Planck scale. This strategy has also been followed for many scenarios for the early universe such as inflation \cite{Maldacena:2002vr}, ekpyrosis \cite{Lehners:2007wc} and the matter bounce \cite{Cai:2011zx}.

It is, therefore, of interest to compute the corrections that self-interactions between perturbations introduce in cosmological observables. If these corrections are large, then the perturbation expansion fails. But even if the perturbation theory is shown to be under control, interactions among perturbations can still be strong enough to generate sizable non-Gaussian correlations in the CMB, while there are strong observational upper bounds \cite{Ade:2015ava}.

A detailed analysis of non-Gaussianity within LQC has been carried out by extending the dressed metric approach to next-to-leading order in perturbations \cite{Agullo:2017eyh}. See \cite{Agullo:2015aba, Zhu_2018,Wu:2018sbr} for preliminary work in this direction, where part of theses non-Gaussianities were discussed. 

There are two quantities of interest: corrections to the power spectrum $\Delta P_{\mathcal{R}}(k)$ induced by next-to-leading order contributions, as well as the size and form of the three-point correlation function $\langle  {\mathcal R}_{\vec k_1}{\mathcal R}_{\vec k_2} {\mathcal R}_{\vec k_3} \rangle$ that is identically zero in the linear (or Gaussian) approximation. We start by reviewing results for the three-point function. 

The three-point functions are quantified by the bispectrum $B_{\mathcal R}(\vec k_1,\vec k_2, \vec k_3)$ defined by $\langle  {\mathcal R}_{\vec k_1}{\mathcal R}_{\vec k_2} {\mathcal R}_{\vec k_3} \rangle=(2\pi)^3\, \delta(\vec k_1+\vec k_2+\vec k_3)\, B_{\mathcal R}(\vec k_1,\vec k_2, \vec k_3)$. The Dirac delta distribution is a consequence of homogeneity, and enforces that $\vec k_1$, $\vec k_2$ and $\vec k_3$ form a triangle. It is convenient to encode the amplitude of the bispectrum in a dimensionless function $f_{\rm NL}(\vec k_1, \vec k_2)$, defined as
\be \label{bisp}
B_{\mathcal R}(\vec k_1,\vec k_2, \vec k_3)= f_{\rm NL}(\vec k_1, \vec k_2) \, \left[\frac{4\pi^4}{k_1^3k_2^3}P_{\mathcal R}(k_1)P_{\mathcal R}( k_2) + \, 
{\rm cyclic\ permutations}\right] \, ,
\ee
where $P_{\mathcal R}( k_i)$ is the power spectrum and the last two terms are obtained from the first one by cyclic permutations of $k_1$, $k_2$ and $k_3$.  
The $f_{\rm NL}(\vec k_1, \vec k_2)$ have been computed numerically in LQC with an post-bounce inflationary phase \cite{Agullo:2017eyh} (assuming a vacuum state for the perturbations far in the contracting branch when all modes of interest were deep inside the curvature radius), with the following results.

First, $f_{\rm NL}(\vec k_1, \vec k_2)$ is highly oscillatory around a small number of the order of the slow-roll parameters (and equal to the value of $f_{\rm NL}$ in standard inflation), see Fig.~2 in \cite{Agullo:2017eyh}. Similar oscillations also appear for the power spectrum, although in that case the oscillations do not average to a small number (see Fig.~\ref{fig:spectra}); these oscillations originate from the oscillatory nature of scalar perturbations.

For large wavenumbers $k\gtrsim k_{\rm LQC}$, $f_{_{\rm NL}}(\vec k_1, \vec k_2)$ reduces to the well-known prediction from slow-roll inflation that $|f_{_{\rm NL}}|\sim 10^{-2}$ \cite{Maldacena:2002vr}, as expected given the discussion at the end of Sec.~\ref{s.inf-app}---this is a good test of the calculations. On the other hand, when the three wavenumbers $k_1$, $k_2$ and $k_3$ are smaller than $k_{\rm LQC}$, then $|f_{_{\rm NL}}|$ grows to $\sim 10^3$. The form of $f_{_{\rm NL}}$ is peaked on wavenumber configurations for which  $k_3 \ll k_2 \approx k_1$,  and  $k_3+k_2 \approx   k_1$; these are called, respectively, squeezed and flattened configurations.

For $k>k_I$, (where $k_I$ is defined as the most infrared mode that exits the horizon during inflation; see Fig.~\ref{fig:spectra}) the modulus of $f_{\rm NL}(\vec k_1, \vec k_2)$ can be approximated by \cite{Agullo:2017eyh}
\begin{equation}\label{eqn:shape}
|f_{\rm NL}(\vec k_1, \vec k_2)|\,\simeq F_{\rm NL}\, e^{-\alpha\,(k_1\,+\, k_2\,+\,k_3)/k_{LQC}}, 
\quad ~
\alpha=\sqrt{\f{\pi}{12}} \, \cdot \frac{\Gamma[5/6]}{\Gamma[4/3]}\approx 0.647,
\end{equation}
where the amplitude $F_{\rm NL}$ is found numerically to be $\sim 10^3$. This shows that $f_{\rm NL}(\vec k_1, \vec k_2)$ is scale-dependent in LQC, with the dependence dictated by $k_{LQC}$; this is a distinctive feature due to the LQC bounce. Like for the power spectrum, LQC effects for non-Gaussianities only become important for wave numbers comparable to (or smaller than) $k_{LQC}$, whose value depends on the number of $e$-folds of inflation. Note that Eq.~\eqref{eqn:shape} is an approximation for the modulus, as it does not include the  oscillations in $f_{\rm NL}(\vec k_1, \vec k_2)$;
an improved approximation which does include the oscillations has been recently introduced \cite{K:2023gsi}. Also, for $k<k_I$ the bispectrum quickly falls off, as also happens for the power spectrum, as can be seen in Fig.~\ref{fig:spectra}; hence $k_I$ can be seen as an infrared cut off.

The predicted amplitude $|f_{\rm NL}(\vec k_1, \vec k_2)|$ is smaller than the upper bounds on non-Gaussianities obtained from  CMB observations \cite{Ade:2015ava}. Furthermore, the recent analysis in \cite{K:2023gsi} shows that even if $k_{LQC}$ lies in the observational window of the CMB, the oscillations of $f_{\rm NL}(\vec k_1, \vec k_2)$ are very effective in washing away the imprint in the CMB bispectrum of the non-Gaussianity produced at the bounce. (This is because the CMB bispectrum is obtained from $f_{\rm NL}(\vec k_1, \vec k_2)$ by smearing out  $\vec k_1$ and $\vec k_2$ against spherical Bessel functions, and the oscillations drastically reduce the result of this smearing.) Hence, it is not expected that such non-Gaussianities will be found in the CBM bispectrum. This is compatible with the recent data analysis performed in \cite{Delgado:2021mxu, vanTent:2022vgy}. However, this does not mean that the primordial perturbations are Gaussian in inflationary models in LQC, but instead it implies that we need to look elsewhere to find its effects. One such possibility could be the large-scale anomalies in the CMB power spectrum, as discussed in Sec.~\ref{s.inf-app}.

Finally, the correction to the power spectrum $\Delta P_{\mathcal{R}}(k)$ at next-to-leading order in perturbations has been shown to be $\Delta P_{\mathcal{R}}/P_{\mathcal{R}}\sim \epsilon\, f_{_{\rm NL}}\, P_{\mathcal{R}}$ \cite{Agullo:2017eyh}, where $\epsilon \lesssim 10^{-2}$ is the slow-roll parameter during inflation. This expression produces $\Delta P_{\mathcal{R}}/P_{\mathcal{R}}\lesssim 10^{-4}$. Therefore, although the amplitude  $f_{_{\rm NL}}$ in LQC is larger by several orders of magnitude than its value in slow-roll inflation, the perturbative expansion remains valid. The smallness of  $\Delta P_{\mathcal{R}}/P_{\mathcal{R}}$ in LQC, even though $| f_{_{\rm NL}}|$ is large, is due to the fact that higher-order perturbative corrections are proportional to higher powers of $P_{\mathcal{R}}(k)$, and $P_{\mathcal{R}}(k)\lesssim 10^{-7}$ is small. In this sense, $P_{\mathcal{R}}(k)$ is the small `parameter' that ensures the validity of the perturbative expansion.

\subsection{Anisotropies}
\label{anisotropies}

Anisotropies are an important topic in bouncing cosmologies, since they rapidly grow in a contracting universe. In general relativity (and in absence of sources of anisotropy), in homogeneous spacetimes the shears are proportional to the inverse cube of the scale factor, so even the tiniest deviation from isotropy would tend to grow rapidly in the contracting phase of the cosmos, more rapidly than contributions from cold dark matter and radiation fluids, and could dominate the dynamics. It is therefore important to check how the predictions for the CMB summarized above (and derived assuming isotropy) may change if some degree of anisotropy is included. This has been analyzed in detail for the simplest anisotropic (though still homogeneous) background spacetime, namely the Bianchi~I geometries \cite{Agullo:2020wur, Agullo:2020iqv, Agullo:2022klq}. 

For the homogeneous background, a study of inflation in anisotropic Bianchi I models within the effective theory of  LQC shows that, first, the attractor character of inflation persists, and second, that solutions of the effective equation which are of phenomenological interest, the universe isotropizes both to the future and past of the bounce \cite{Gupt:2013swa}.  This shows that the  so-called ``cosmic no-hair theorem'' of general relativity \cite{Wald:1983ky}, that anisotropies in the early universe are generically washed out by the expansion, remains true in LQC.

On the other hand, cosmological perturbations retain memory of the anisotropies for  much longer than the  background geometry does \cite{Agullo:2020wur, Agullo:2020iqv, Agullo:2022klq}. This is because anisotropic features in the perturbations do not dilute as the universe expands, although they are red-shifted in the sense that wavenumbers $\vec k$ of the modes with non-zero anisotropies can be red-shifted to super-Hubble scales and made inaccessible to observations. Since the red-shift is linear with the scale factor, unless the accumulated expansion in the cosmic history is much larger than what it is commonly accepted, perturbations can evade the cosmic no-hair theorem and imprint some degree of anisotropy in the CMB, even though anisotropies in the background geometry may be completely diluted by the expansion. 

The  theory of comological perturbations on a background Bianchi~I geometries is significantly more complicated than its counterpart on FLRW spacetimes. The equations of motion for scalar and tensor perturbations can be derived for classical general relativity by expanding Einstein's equations \cite{Pereira:2007yy}, as well as in a Hamiltonian treatment, better adapted to LQC, where the quantization can also be completed \cite{Agullo:2020uii, Agullo:2020kil}. Anisotropies introduce two main new features for the dynamics of the perturbations, by modifying the effective potentials in the perturbative equations of motion such that they: (i) now depend on the direction of the wavenumber $\vec k$ and, (ii) couple scalar and tensor modes. These potentials induce anisotropies in the perturbations, even if their initial state is isotropic, as well as correlations---and quantum entanglement---among scalar and tensor perturbations. In particular, the scalar and tensor power spectra of the isotropic theory are replaced by a family of direction-dependent power spectra ${P}_{ss'}(\vec k)$, with $s,s'=0,+2,-2$, where $s=\pm 2$ describe circularly polarized tensor modes. Anisotropies in ${P}_{ss'}(\vec k)$ can be quantified by their angular multipoles, obtained by expanding the spectra in spin-weighted spherical harmonics
\be \label{PssLM}
{ P}_{ss'}(\vec k) = \sum_{L=|s-s'|}^{\infty} \sum_{M=-L}^L \,
{ P}^{LM}_{ss'}(k)\ _{s-s'} Y_{LM}(\hat k) \, ,
\ee
where $_{s-s'}Y_{LM}(\hat k)$ is a spherical harmonic of weight $s-s'$. The use of spin-weighted  spherical harmonics guarantees that all ${ P}^{LM}_{ss'}(k)$ are scalars under rotations. The coefficients ${ P}^{LM}_{ss'}(k)$, which depend only on the modulus of $\vec k$, encode the information  about anisotropies in the primordial perturbations, and are all zero in the isotropic limit except for $L=M=0$. In Bianchi~I geometries, $L$ is constrained to take even values, if the initial state for perturbations is parity invariant. Multipolar components ${ P}^{LM}_{ss'}(k)$ with $s\neq s'$ describe correlations between different perturbations. 

The  functions ${P}^{LM}_{ss'}(k)$ have been computed in LQC for inflationary models \cite{Agullo:2020wur, Agullo:2020iqv} as well as for the ekpyrotic and matter bounce scenarios \cite{Agullo:2022klq}, with the common prediction in all these models that anisotropies in the primordial spectra are dominated by the quadrupolar contribution $L=2$. The key difference between different models is the predicted scale-dependence of the quadrupole: for inflationary models, ${ P}^{2M}_{ss'}(k)$ scales approximately as $1/k$, while for ekpyrosis and the matter bounce ${ P}^{2M}_{ss'}(k)$ is almost scale invariant. A quadrupolar modulation in the CMB is the hallmark of primordial anisotropies, and its scale-dependence can be used to distinguish between inflation and some of its alternatives.

Interestingly, a quadrupolar modulation has been detected in the CMB \cite{Planck:2018jri}, although its statistical significance is low, since the chances that such quadrupole could have been generated in an isotropic universe as a result of a statistical fluke are high. Nonetheless, the observed quadrupole may have a natural explanation in primordial anisotropies generated by a cosmic bounce \cite{Agullo:2020wur, Agullo:2020iqv, Agullo:2022klq}. Future observations, particularly if tensor modes are detected, will help to clarify the origin of the quadrupole and its scale dependence. 

It is also possible to compute angular correlation functions in the CMB, $C_{\ell\ell'}^{XX'}$ where $X,X'=T,E,B$ refers to temperature, electric, and magnetic polarization in the CMB, including their cross-correlations, in LQC \cite{Agullo:2020iqv, Agullo:2022klq}. In particular $TB$ and $EB$ correlations are forbidden by symmetry arguments in FLRW spacetimes, while they are non-zero in Bianchi~I and are therefore a smoking gun for anisotropies. Finally, Ref.~\cite{Agullo:2020iqv} also contains a Markov chain Monte Carlo analysis (MCMC), using TT, EE, TE, and lensing data to find the best fit to the six free cosmological parameters $\Omega_b$, $\Omega_c$, $\theta_{MC}$, $\tau$, $A_s$ and $n_s$ in presence of anisotropies.

\section{Limitations}
\label{s.lim}

LQC has been able to provide a detailed possible picture for the Planck era of the universe. This scenario has been applied to extend existing viable cosmological models based on general relativity to include Planck-scale physics. The inclusion of cosmological perturbations described in this chapter has lead to a better understanding of the way Planck-scale physics could be imprinted in the  observables we have access to at present, mainly through the CMB. 

The main limitations of this research program are rooted in the absence of a complete theory of quantum gravity. This is what motivated the development of LQC at first, as a symmetry reduced version of LQG. These limitations are carried over to the description of perturbations. The strategy followed so far for LQC and cosmological perturbations is a common one in physics: start from the simplest scenario and add complications in a sequential manner, in order to test the robustness of the predictions. One expects that, although fine details may change as one builds more complete models, the main features will be robust. This strategy has been very successful in the history of physics, from the  study of the hydrogen atom to black holes. Regarding perturbations in LQC, this was the motivation to study anisotropies and non-Gaussianity. But there still remain several avenues where the robustness of the theoretical framework needs to be further tested. We summarize in this section the status of the main two limitations, namely the absence of a loop quantization for cosmological perturbations and the existence of quantization ambiguities.

\subsection{Trans-Planckian modes}
\label{s.tp}

In LQG, the simplest geometric observable is the two-dimensional area of a surface. The spectrum of the area operator is discrete and has the important property that there is a minimum non-zero area eigenvalue $\Delta \sim \lp^2$, called the area gap. In this sense, there is an ultraviolet cutoff for the area in LQG. This raises the obvious question: is there an ultraviolet cutoff for the wavelength of cosmological perturbations? In other words, do there exist cosmological perturbations with a trans-Planckian wavelength $\lambda < \lp$, or not? This is the trans-Planckian problem in cosmology.

Note that the trans-Planckian problem is of particular interest for inflationary models, since if the inflationary period is sufficiently long then some of the CMB modes would have been trans-Planckian at the onset of inflation, in which case the trans-Planckian problem becomes observationally relevant. But even for alternatives to inflation like ekpyrosis or the matter bounce, the trans-Planckian problem is still present as a conceptual problem even though it may not be observationally relevant.

The possibility of an ultraviolet cutoff for cosmological perturbations, motivated by quantum gravity, has been proposed in various contexts \cite{Weiss:1985vw, Jacobson:1999zk, Martin:2000xs, Kempf:2001fa}. On the one hand, it could cure ultraviolet divergences in the quantum theory, but on the other hand it seems to require a dynamical number of degrees of freedom since the ultraviolet cutoff is $k < 1 / a(t) \lp$ (although this would presumably not be a problem for full LQG where the Hilbert space includes all possible graphs). A na\"ive cutoff can also induce unwanted violations of local Lorentz invariance: even if the cutoff is at the Planck scale, such a violation can produce prohibitively large effects at low energy due to the integrals in $k$ appearing in radiative corrections \cite{Collins:2004bp}. 

While the presence of an ultraviolet cutoff for the LQG area observable is suggestive that there may also be a cutoff for the wavelengths of cosmological perturbations in LQC, it is not conclusive. There are several proposals for length operators in LQG \cite{Thiemann:1996at, Bianchi:2008es, Ma:2010fy}, and while they have a discrete spectrum, numerics suggest eigenvalues can be arbitrarily close to 0 \cite{Brunneman:2007as}. This discussion is quite preliminary, and more work is needed to determine whether there is or not an ultraviolet cutoff for the length operator, and the implications for cosmological perturbations.

In the dressed metric approach, the adiabatically-renormalized energy and pressure densities of the perturbations always remain below the Planck scale, even near the bounce \cite{Agullo:2013ai}. Since LQC effects only become important for the background dynamics when the energy density approaches the Planck scale, this (combined with the absence of an obvious minimum length in LQG) suggests that it may be sufficient to concentrate on LQC effects on the background degrees of freedom and to assume that LQC effects in perturbations are subleading, in particular for the longest wavelengths we observe in the CMB. This motivates the Fock quantization of perturbations in both the dressed metric and hybrid approaches. But it is a strong assumption, and it is desirable to reinforce it from the point of view of full LQG. 

For a phenomenological study of potential trans-Planckian effects in LQC, see \cite{Martineau:2017tdx} where modified dispersion relations motivated by analog gravity \cite{PhysRevD.51.2827} were considered; the result is that if there are sufficiently many $e$-folds of inflation, in some cases these modified dispersion relations can have an impact on predictions for the CMB. It remains to be seen whether modified dispersion relations can arise due to LQC effects on cosmological perturbations, and if so what specific modified dispersion relation captures the LQC corrections.

It would be nice to study this question using a loop quantization of perturbations, like the approach based on the separate universe framework \cite{Wilson-Ewing:2012dcf} summarized above, but this approach is limited to long-wavelength modes and consequently it cannot shed any insight on the trans-Planckian problem; see Sec.~\ref{s.sep} for details. Instead, it may be necessary to go beyond LQC and study cosmology from full LQG; for recent work in this direction see Sec.~\ref{s.lqg}.

In summary, whether the possibility of a Planckian cutoff is realized or not in LQC, it is clearly important to understand the impact of LQG effects on trans-Planckian modes: are they ruled out by LQG, or do they exist with possibly modified dynamics? More generally, recall that for FLRW spacetimes, LQC effects (as expressed for sharply-peaked states that can be well approximated by an effective description) modify the classical equations of motion with terms of the order $\rho / \rho_{\rm Pl}$, and it is well understood that corrections of this type affect cosmological perturbations as well. For Fourier modes of perturbations whose wavelength nears (or perhaps  is even shorter than) $\lp$, it is possible that LQG effects may modify their dynamics with extra quantum corrections of the form $\lambda / \lp$, with the exact form of these corrections remaining to be determined. These are important open questions that could lead to a much better understanding of quantum gravity effects in cosmology.

\subsection{Quantization ambiguities}

A second problem in LQC (as well as in LQG or, more generally, in non-linear quantum theories) is the presence of ambiguities that arise in the quantization process. Generally speaking, there are two types of quantization ambiguities that are relevant for cosmological perturbations: ambiguities arising in the loop quantization of the FLRW background, and ambiguities concerning the perturbations themselves.

There are several sources of quantization ambiguities for the loop quantization of the FLRW spacetime, and these ambiguities will of course affect the dynamics of perturbations evolving on this background. These include ambiguities in the definition of the curvature and inverse volume operators \cite{Singh:2013ava}, as well as the ambiguities related to modified LQC that have been discussed above in Sec.~\ref{s.modlqc}, specifically which form of the Hamiltonian constraint to quantize and whether it is necessary to use Thiemann's identity to express the extrinsic curvature as an operator. The first category of ambiguities is less important as it leads to some quantitative differences in the quantum evolution, but qualitatively the dynamics are not significantly affected by these ambiguities, at least for the flat FLRW spacetime. On the other hand, the ambiguities of modified LQC produces important differences in the quantum dynamics of the background spacetime in the contracting branch before the bounce, and can have an impact on the perturbations evolving on this background, as explained in Sec.~\ref{s.modlqc}. For this reason, it is important to better understand how to address these ambiguities properly.

Another ambiguity is closely related to the problem of time. In LQC, the problem of time is usually addressed by using a matter field as a relational clock, so that the quantum evolution of the cosmological wave function is calculated with respect to the matter field. While this is a natural way to address the problem of time in this context, in general there may be multiple matter fields present and then there is an ambiguity: which matter field should be used as a relational clock? Further, it is also possible to use geometric clocks as well as matter clocks \cite{Martin-Benito:2008dfr, Giesel:2018opa}, which clearly increases the number of possible clocks. Presumably, it will always be possible to choose different clocks, but (although this point is clear in the classical theory) relating the quantum evolution with respect to different relational clocks is not always simple \cite{Vanrietvelde:2018pgb, Gielen:2020abd}.

Finally, the last ambiguity is due to the absence of a complete loop quantization for cosmological perturbations. There are four main approaches that have been developed, as reviewed above: the dressed metric approach, hybrid quantization, the separate universe framework for LQC, and effective dynamics. All of these approaches have their strengths, although none can be considered complete. At this time, it is necessary to make a choice on which approach to use (although the dressed metric and hybrid approaches give very similar results \cite{Li:2022evi}). As discussed at the beginning of this section, an important open problem is to understand how to perform a loop quantization for all cosmological perturbations (and which will hopefully give these four approaches in various limits). Note that clarifying the connection to full LQG may help both with resolving these ambiguities as well as developing a loop quantization for the perturbations.

\section{Beyond LQC}
\label{s.lqg}

LQC is based upon applying the quantization tools of LQG to cosmological spacetimes. But LQC is a symmetry-reduced model, in which the symmetries of the spacetimes of interest are imposed at the classical level, before quantization. The processes of quantization and symmetry reduction do not  commute in general, so there is no guarantee that LQC does not miss some important aspects of LQG. Symmetry reduction has been  useful to make progress from black holes to atomic physics, but it is clearly desirable to eventually connect LQC with full LQG. Further, since there are ambiguities in the definition of LQG itself, connecting LQC to LQG may also give some suggestions on how to address the ambiguities of LQG. There have been some promising results---for both background and perturbative degrees of freedom---coming from three different directions: canonical LQG, covariant spin foam models, and group field theory.

For the homogeneous and isotropic FLRW spacetime, there has been significant progress in extracting cosmology from full LQG. To do this requires several non-trivial steps. The first step is to pick a certain approach to LQG and a particular definition of the dynamics; given the quantization ambiguities in LQG, it is interesting to explore different possibilities and determine whether some choices are preferred over others. The second step is to choose a certain class of quantum states that may correspond to cosmological spacetimes; this is done by determining how to appropriately impose the conditions of homogeneity and isotropy on quantum states (note that these conditions cannot be imposed exactly due to quantum fluctuations). The third step is conceptually simple but often technically difficult, which is to evaluate the quantum dynamics, as prescribed by step 1, on the states chosen in step 2. In general, approximations are often needed in this step since the quantum dynamics typically is not tangential to the subspace of cosmological states chosen in step 2.

For canonical LQG, there has been work relating LQC to full LQG at both the kinematical and dynamical levels. Kinematically, it has been shown how to define a diffeomorphism-invariant notion of a homogeneous and isotropic sector of LQG and how to embed the LQC kinematical Hilbert space into the one of LQG \cite{Brunnemann:2007du, Fleischhack:2010zt, Ashtekar:2012cm, Beetle:2016brg, Engle:2016hei, Beetle:2017qle, Engle:2016zac, Engle:2018zbe, Engle:2019zfp} (see \cite{Beetle:2021zsv} for an extension to Bianchi~I geometries). Multiple approaches have been developed at the dynamical level; three based on studying coherent states on a fixed graph structure in LQG are: (i) the quantum-reduced loop gravity approach where the additional symmetry that the metric be diagonal is imposed on the quantum theory as a weak constraint, reducing the SU$(2)$ labels on the graph to U$(1)$ quantum numbers \cite{Alesci:2013xd, Alesci:2016gub}; (ii) using complexifier coherent states and treating the Euclidean and Lorentzian terms in the scalar constraint separately as in full LQG, which is related to the modified LQC theory mLQC-I (see Sec.~\ref{s.modlqc}) \cite{Dapor:2017rwv,Dapor:2017gdk}; and (iii) using a path-integral reformulation of LQG \cite{Han:2019vpw} as well as a perturbative expansion of the Hamiltonian constraint \cite{Zhang:2021qul}; these three models also give results similar in some ways to LQC or modified LQC, although with some important drawbacks concerning the classical limit \cite{Pawlowski:2014nfa, Dapor:2019mil} and not matching some aspects of the LQC/mLQC-I dynamics \cite{Ashtekar:2006wn, Yang:2009fp} known in the LQC literature as the `improved dynamics'. To address these shortcomings, the above work has since been extended to allow for graph-changing dynamics for the approach based on the path-integral reformulation of LQG \cite{Han:2021cwb}, while other approaches have also been developed with a key ingredient being length-dependent holonomies \cite{Bodendorfer:2016tky, Han:2019feb}. Both of these approaches, with graph-changing or length-dependent holonomies, give the correct classical limit. Further, the dynamics in the Planck regime are very similar to mLQC-I for the graph-changing LQG dynamics \cite{Han:2021cwb}, while the approach based on length-dependent holonomies gives dynamics very similar to standard LQC \cite{Bodendorfer:2016tky, Han:2019feb}.

For the covariant spin foam approach, the dynamics is based on the spin foam vertex amplitude for vertices of arbitrary valence defined in \cite{Kaminski:2009fm}, while the quantum states are coarse triangulations of a 3-sphere, capturing large-scale degrees of freedom \cite{Bianchi:2010zs, Bianchi:2011ym, Rennert:2013pfa}. It seems likely necessary to extend these models to either allow for graph-changing or length-dependent holonomies, as in the canonical case, to recover the correct dynamics. But already some important qualitative similarities to LQC arise, like a non-singular bounce, and suggest the connection between LQG and LQC will be possible in spin foam models as well.

For group field theory (GFT), the dynamics is based on the quantum equations of motion for the GFT field, and the quantum states corresponding to FLRW spacetimes are assumed to be condensate states, where homogeneity is imposed by assuming every quantum of geometry is in the state \cite{Gielen:2013kla}; this approach is based on viewing cosmology as the hydrodynamical limit of GFT \cite{Oriti:2010hg}. Using a massless scalar field as a relational clock, the dynamics can be extracted in a relational form, giving dynamics very similar (and in fact identical for a specific class of states) to the LQC dynamics in the limit that the non-linear term in the dynamics is negligible \cite{Oriti:2016qtz}. For further work on the cosmological sector of GFT, see \cite{Gielen:2016uft, deCesare:2016rsf, Wilson-Ewing:2018mrp, Pithis:2019tvp}.

There has also been work in studying cosmological perturbations starting from full LQG. There was some early work based on the quantum-reduced loop gravity approach \cite{Olmedo:2018ohq, Han:2020iwk}, and importantly this has since been extended to use length-dependent holonomies \cite{Han:2021cwb}. Also, quantum correlations between different spatial regions have been studied in spin foam models \cite{Gozzini:2019nbo} and two approaches to perturbations in GFT have been studied, one based on using matter fields as reference rods to localize perturbations \cite{Gielen:2017eco, Marchetti:2021gcv}, and the other on an extension of the separate universe approach for long-wavelength scalar perturbations to GFT states \cite{Gerhardt:2018byq}.

The progress made so far in understanding the relation between LQC and LQG is encouraging, but remains incomplete. The three approaches developed so far each have their strengths and weaknesses. Canonical LQG offers a relatively direct path to recover cosmological dynamics, but it is difficult to handle graph-changing dynamics which have been argued to be important. It may be possible to sidestep this problem by using length-dependent holonomies, but this is a new ingredient that would also need to be implemented in full LQG. The work in the spin foam approach suggests using especially simple quantum states to describe cosmological spacetimes, but it seems that more complicated states (i.e., more refined triangulations of a 3-sphere) will be needed for cosmological perturbation theory, especially to describe short-wavelength perturbations. And for GFT, condensate states seem to capture the notion of homogeneous states extremely well, with their dynamics very similar to LQC, but these condensate states ignore the graph structure of the quantum state and this may lead to a too-rapid growth in the strength of the non-linear term in the GFT dynamics.

It is our hope that future work in these various directions will address these challenges, and further improve our understanding of quantum gravity effects in cosmology, both at the background and perturbative levels.

\medskip

\noindent
{\it Acknowledgments:}

I.A.~is supported by the NSF grant PHY-2110273, and by the Hearne Institute for Theoretical Physics.
E.W.-E.~is supported by the Natural Sciences and Engineering Research Council of Canada, and by the UNB Fritz Grein Research Award.
We thank S. Ducoing for assistance with Fig.~\ref{fig:diagram}.

\end{document}